# Antiferromagnetic order in the Hubbard Model on the Penrose Lattice

Akihisa Koga
*Department of Physics, Tokyo Institute of Technology, Meguro, Tokyo 152-8551, Japan*

Hirokazu Tsunetsugu
*The Institute for Solid State Physics, The University of Tokyo, Kashiwa, Chiba 277-8581, Japan*

We study an antiferromagnetic order in the ground state of the half-filled Hubbard model on the Penrose lattice and investigate the effects of quasiperiodic lattice structure. In the limit of infinitesimal Coulomb repulsion $U \to +0$, the staggered magnetizations persist to be finite, and their values are determined by confined states, which are strictly localized with thermodynamics degeneracy. The magnetizations exhibit an exotic spatial pattern, and have the same sign in each of cluster regions, the size of which ranges from 31 sites to infinity. With increasing $U$, they continuously evolve to those of the corresponding spin model in the $U = \infty$ limit. In both limits of $U$, local magnetizations exhibit a fairly intricate spatial pattern that reflects the quasiperiodic structure, but the pattern differs between the two limits. We have analyzed this pattern change by a mode analysis by the singular value decomposition method for the fractal-like magnetization pattern projected into the perpendicular space.

## I. INTRODUCTION

Recently, electron correlations in quasiperiodic systems have attracted much interest since the report of quantum critical behavior in the quasicrystal $Au_{51}Al_{34}Yb_{15}$ [1]. Investigation of the interplay of the two effects, strong correlation and lattice quasiperiodicity, is a big challenge, since each of the two is already a big issue. Therefore, to understand generic important characteristics in their interplay, it is useful to employ simpler quasiperiodic lattices such as the one-dimensional Fibonacci lattice [2–5] and two-dimensional octagonal lattice [6–8]. One of the simplest ones with geometrical quasiperiodicity is the Penrose lattice in two dimensions [9–17]. On this lattice, a few strongly correlated models have been studied: the Ising model for classical spins [18–20], the Heisenberg model for quantum spins [21, 22], and the Hubbard [23, 24] and Anderson lattice models [25, 26] for correlated electrons. One important feature of the Penrose lattice is the presence of thermodynamically degenerate one-particle states, and they are called *confined states*, since the their wave functions are strictly confined in a finite region in space [9, 14]. One of the main issues in this paper is how these states respond to electron interactions.

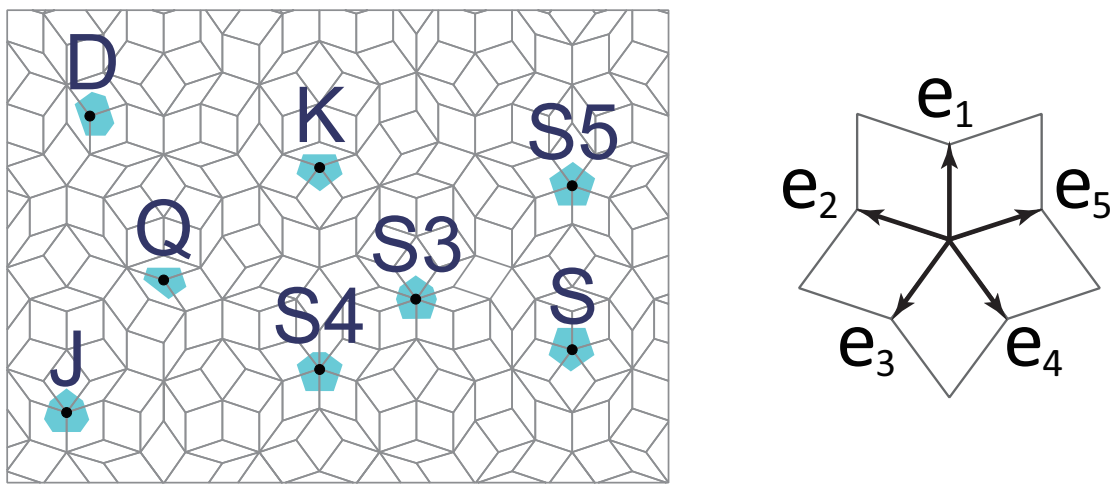

FIG. 1: (Color online) Penrose lattice and eight types of vertices with nomenclature of de Bruijn [27, 28]. The shaded regions represent the corresponding Voronoi cells. $\mathbf{e}_1, \cdots, \mathbf{e}_5$ are projection of the fundamental translation vectors in five dimensions, $\mathbf{n}=(1,0,0,0,0), \cdots, (0,0,0,0,1)$.

In this paper, we will study the Hubbard model on the Penrose lattice. As for the atomic sites, we employ the so-called vertex model: electrons locate at vertices of the rhombus units and hop along their edges. See Fig. 1. In this case, the lattice is bipartite and this implies a symmetric energy spectrum of electrons. The Fermi level is fixed at $E_F = 0$ at half filling, due to the particle-hole symmetry of the Hamiltonian, even when electron interactions are switched on. In the vertex model, all the confined states appear at the energy $E = 0 = E_F$, and therefore one expects that they have important contributions in many properties. We will show that their thermodynamic degeneracy particularly plays a key role in formation of magnetic order in the weak coupling region. This situation may seem similar to that in the flat band ferromagnetism [29–31], but the stabilized order is an antiferromagnetic one in the present case.

In order to analyze contributions of the confined states, one needs detailed information on their number and wave function amplitudes defined in such a way that different states are mutually orthogonal, which is a standard prerequisite for the Hamiltonian's basis states. Regarding the number counting, one also needs to group confined states into various types, and one problem is whether all the possible types have already been discovered. In this paper, we also solve these problems.

In the lattice, confined states are separated by stripe regions which were named *forbidden ladders* in Ref. [14]. Using this property, we will introduce useful structure units, *clusters*, for defining the domain of confined states, and then show a few exact results about the confined states. We will prove the Arai's conjecture [14] that the grouping is completed with six types of confined states. Because of their degeneracy, each confined state is not uniquely defined, but a natural choice should be such that each state has finite amplitudes only in one of the two sublattices. Examining the profiles of clusters in detail, we also show that the two sublattices have the same number of confined states in the thermodynamic limit. This confirms that the uniform magnetization density vanishes in the antiferromagnetic order, and this excludes the possibility of ferrimagnetic orders. Furthermore, we obtain the distribution of local magnetizations in the weak coupling limit. We will also discuss the crossover to the strong coupling limit within the Hartree-Fock approach.

The paper is organized as follows. In Sec. II, we introduce the Hubbard model on the Penrose lattice. In Sec. III, we define the clusters in the Penrose lattice, examining the confined states. Then, we obtain some exact results and discuss magnetic properties in the weak coupling limit in Sec. IV. Using the Hartree-Fock approximation, we discuss the crossover between weak coupling and strong coupling states in the antiferromagnetically ordered phase in Sec. V. Exotic spatial pattern is a characteristic feature of the antiferromagnetic order in this system, and it is analysized in detail in Sec. VI A summary is given in the last section.

## II. MODEL

The Hamiltonian to study in this paper is the half-filled Hubbard model with sites on the vertices of the Penrose lattice (see Fig. 1)

$$H = -t \sum_{\langle j,j'\rangle\sigma} c^{\dagger}_{j\sigma} c_{j'\sigma} + U \sum_{j} (n_{j\uparrow} - \tfrac{1}{2})(n_{j\downarrow} - \tfrac{1}{2}), \qquad (1)$$

and we will study the antiferromagnetic order in its ground state. Here the sum of $\langle j, j'\rangle$ should be taken over nearest neighbor pairs of sites, $c_{j\sigma}$ is the electron annihilation operator at the site $j$ with spin $\sigma \in \{\uparrow,\downarrow\}$ and $n_{j\sigma} \equiv c^{\dagger}_{j\sigma} c_{j\sigma}$. Throughout this paper, we use $N$ to denote the number of sites. $t$ is the transfer integral of electron hopping, and $U \geq 0$ the Coulomb repulsion of electron interaction. Since the sites are located on the vertices of even-member rings, the lattice is bipartite and the chemical potential is $\mu = 0$ at half filling electron density $n \equiv \frac{1}{N}\sum_{j\sigma}\langle n_{j\sigma}\rangle = 1$. For any value of $U$, the chemical potential does not shift from $\mu = 0$ and this holds also for finite clusters with any shape.

Lieb's theorem [32] for the half-filled Hubbard model states that the ground state on a bipartite lattice has the total spin $S_{\rm tot} = \frac{1}{2}\Delta N_{AB}$, where $\Delta N_{AB} \equiv |N_A - N_B|$ with $N_A$ and $N_B$ being the numbers of sites in the two sublattices. Therefore, the imbalance of site number between the sublattices yields the ferrimagnetically ordered state like in the Lieb lattice. This may also apply to our model, and it is not trivial that $\Delta N_{AB}/N \to 0$ in the thermodynamic limit. In the strong coupling limit $U \to \infty$, the system is reduced to the Heisenberg model with the exchange coupling $J = 4t^2/U$, where the ground state has an antiferromagnetic long-range order [21, 22]. In contrast, interesting behavior should appear in the weak coupling region, since confined states exist at the energy of the chemical potential $E = 0$. Therefore, switching $U$ induces a magnetic order in which the ordered moments are mainly spin polarization of the confined states. To discuss the weak coupling region and crossover to strong coupling, we examine in the next section the confined states in detail with a special attention about their mutual independence, and also introduce a new unit structure. Then, we discuss magnetic properties in the system.

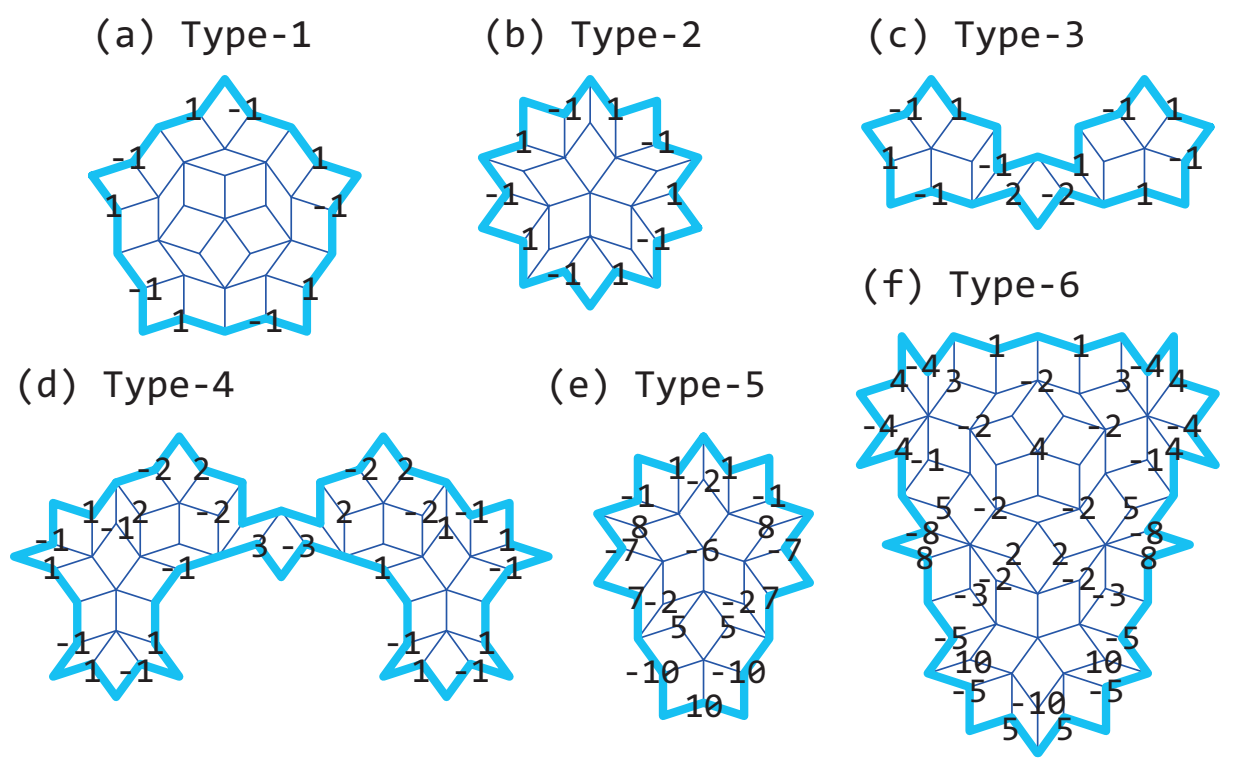

FIG. 2: (Color online) Six types of the confined states in the tight-binding model on the vertices of the Penrose lattice. The numbers at the vertices represent the amplitudes of confined states [9, 14].

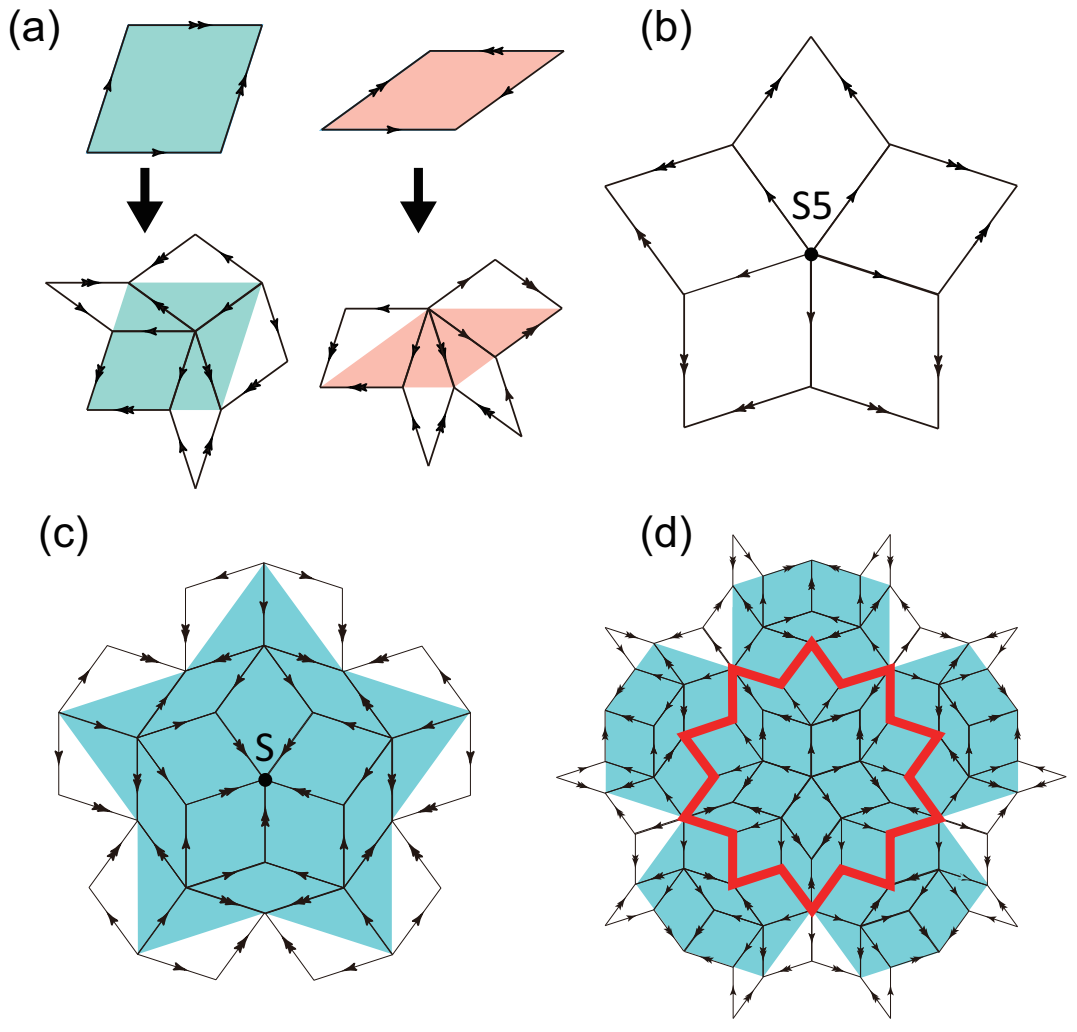

FIG. 3: (a) Deflation rule. (b) Initial pattern centering around an S5-vertex. Patterns after (c) one deflation and (d) two deflations, and the area in the previous step is colored in blue. Bold line in (d) indicates the area of the cluster-1.

## III. CONFINED STATES AND CLUSTERS

The confined states in the noninteracting system are one of the most characteristic properties in the electronic states on the Penrose lattice and have been studied in detail [9, 14]. Arai et al. conjectured that all the confined states are classified to the six types depicted in Fig. 2. Since each confined state satisfies the Schrödinger equation $H(U = 0)\psi = E\psi$ with $E = 0$, it is always possible to choose each eigenstate such that it has finite amplitudes only in one of the two sublattices. Here, we define the domain of each type of confined state by the following two rules: the confined state has zero and a finite amplitudes alternately along its boundary, and its area is maximized. This definition modifies the boundaries of type-3, 4, 5 and 6 confined states slightly from those used in Ref. [14].

Although the Penrose lattice does not have the translation symmetry, its quasiperiodicity is closely related to a special dilatation operation called *deflation*, and the Penrose lattice can be generated by its successive operations starting from any seed pattern. Each deflation transforms rhombus units as shown in Fig. 3 (a), and should be followed by magnification to restore the original rhombus sizes. This magnification ratio is the golden ratio $\tau = (\sqrt{5}+1)/2$, and it satisfies the well-known relation $\tau^2 = \tau + 1$, which indicates a self-similarity of the lattice.

An important property of the Penrose lattice is stated by Conway's theorem [27, 33]. Any finite part of the lattice repeats itself within the distance shorter than twice its diameter. This means that despite the lack of genuine periodicity the Penrose lattice is quite uniform, and this guarantees a finite density of any finite-size pattern in the thermodynamic limit. Since confined states locate in finite-size domains, their densities are also finite, and this leads to a delta-function peak at $E = 0$ in the electron density of states (DOS).

To examine the spatial distribution of the confined states, we cover the lattice with the six types of domains, and the result is shown in Fig. 4. Multiple domains may overlap to cover some larger parts, and these parts are separated by stripes in white color. These stripes are a set of unbranched closed loops made of the rhombuses that do not belong to any confined-state domain. Following Ref. [14], we will refer to these stripes as *forbidden ladders*. We will show that it is useful to introduce a new unit of structure: *clusters* are defined as those connected parts separated by forbidden ladders.

In each cluster, confined states have finite amplitudes only in one of the A/B-sublattices, and their number is given by the imbalance of the sublattice sites $\Delta N^{AB}$. The sublattice hosting confined states will be referred to as *M sites*, while the other sublattice will be called *D sites*. One should note that not all the M sites have finite amplitudes of confined states. Each cluster is repeated quite regularly in the lattice, and depending its position the M sites are either the A-sublattice or the B-sublattice.

Let us examine the relation of the M/D sites and the sublattices. Consider the boundary of one cluster, and the M and D sites are alternatively arranged there. Arai et al. showed that only D sites are connected to neighboring clusters by bonds inside the intervening forbidden ladder [14]. See Fig. 4 (b). Applying the same argument to the neighboring cluster, one finds that both ends of any bond inside the forbidden ladder should be D sites. This means that the M sites changes the A/B-sublattice when one moves from one cluster to its neighboring clusters. This property will turn out important when we study an antiferromagnetic order later. This result also indicates that it is more convenient to examine characteristics of the confined states in terms of these clusters, rather than the classification of vertices [27, 28].

We now analyze in detail spatial profiles of different clusters. See Fig. 4. The smallest one is denoted as *the cluster-1* and made of ten fat and ten skinny rhombuses centering around an S5-vertex. One should also note that applying deflation twice transforms a single S5-vertex (more precisely, five fat rhombuses around an S5-vertex) to a cluster-1 encircled by a forbidden ladder plus additional 20 rhombuses, as shown in Fig. 3.

We find that large clusters are successively generated from the cluster-1 by repeating deflation, and we refer as *the cluster-i* to the one generated by $(i-1)$-times deflations. One needs to notice that forbidden ladders are also be generated, and the outermost ladder and its exterior part should be subtracted in each operation. When other forbidden ladders appear inside, one should also subtract them and their interior parts, and this makes the clusters multiply connected for $i \geq 3$. The property explained above implies that all the S5-vertices in the cluster-$i$ evolve upon double deflations into cluster-1's, which should be subtracted from the cluster-$(i+2)$. We can prove that this is the only necessary subtraction process in the genesis of large clusters. This is because if a larger cluster-$i$ ($i \geq 2$) were to be subtracted, its ancestor cluster-1 would appear already in one of previous deflations and it should have been subtracted at the stage of cluster-1. See also Appendix A. Thus a recursion relation is obtained for the cluster numbers:

$$C^{(l+1)}_{i,s} = \begin{cases} (S5)^{(l-1)}_{-s} & (i=1) \\ C^{(l)}_{i-1,s} & (i \geq 2) \end{cases}. \qquad (2)$$

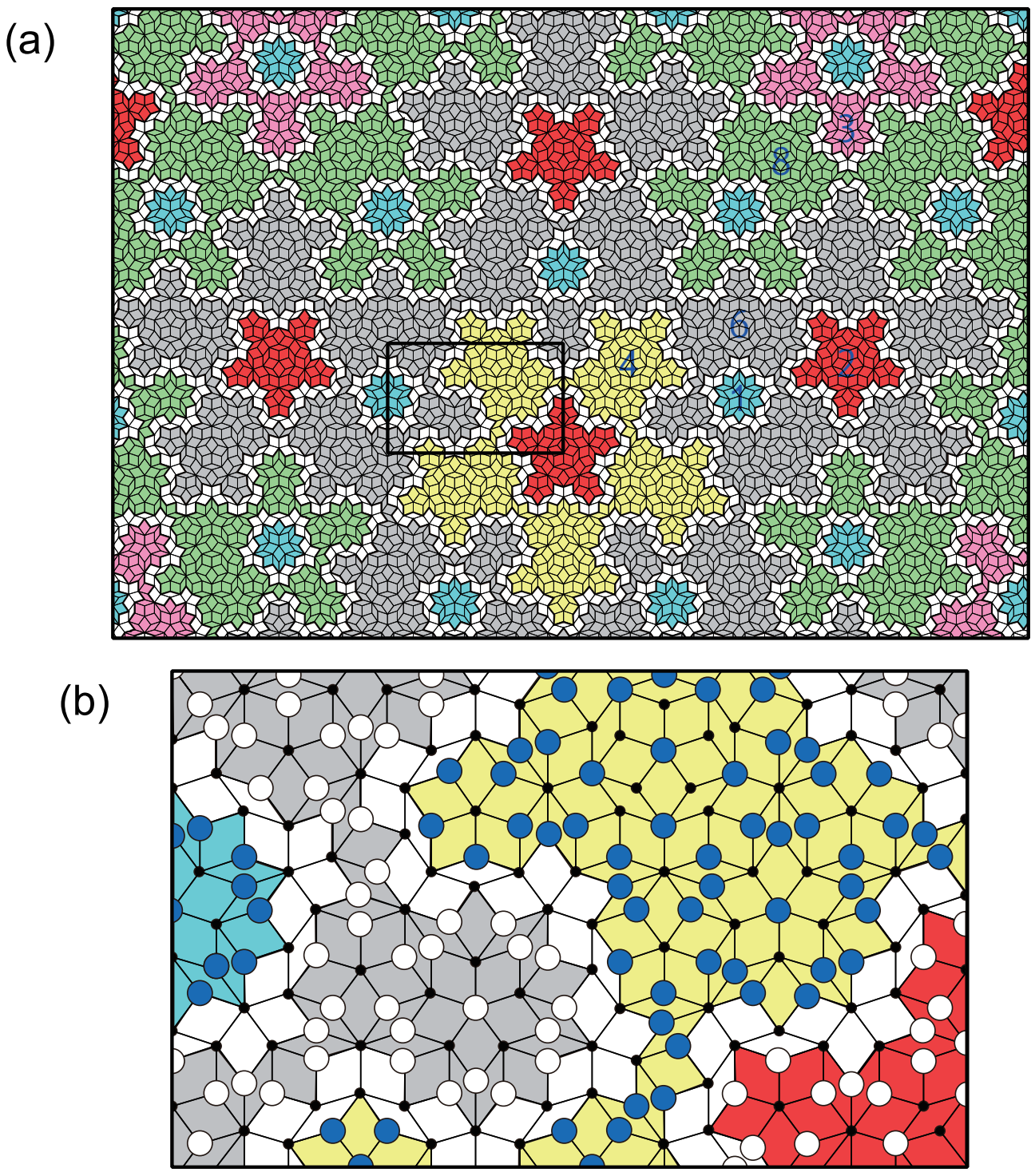

FIG. 4: (Color online) (a) Clusters in the Penrose lattice. The clusters-$i$ are colored in ($i$=1) light blue, (2) red, (3) pink, (4) yellow, (6) gray, and (8) green. White regions are forbidden ladders (the S or S5 strings [14]). (b) Magnified image of the rectangular part in (a). Big circles show the M sites, where confined states may have finite amplitudes. Blue circles are A-sublattice sites, while white ones are in the B-sublattice.

Here, $C_{n,s}$ denotes the number of the cluster-$n$'s with the M sites in the sublattice $s \in \{A, B\}$, and $(S5)_{-s}$ is the S5-vertex number in the sublattice complement to $s$. The superscript $(l)$ means that these numbers are those in the $l$-th stage in repeating deflation. The simplest choice of initial condition is

TABLE I: Cluster profile of each cluster-$i$. Its fraction is $p_i$, the site number $N_i$ and that in each sublattice $N_i^\alpha$ ($\alpha = M, D$), their difference $\Delta N_i^{AB} \equiv N_i^M - N_i^D$; $N_{i,k}$ is the number of type-$k$ confined states, and $(S5)_i$ is the number of S5-vertices.

| $i$ | $\sqrt{5}\,p_i$ | $N_i$ | $N_i^M$ | $N_i^D$ | $\Delta N_i^{AB}$ | $N_{i,1}$ | $N_{i,2}$ | $N_{i,3}$ | $N_{i,4}$ | $N_{i,5}$ | $N_{i,6}$ | $(S5)_i$ |
|---|---|---|---|---|---|---|---|---|---|---|---|---|
| 1 | $\tau^{-11}$ | 31 | 16 | 15 | 1 | 0 | 1 | 0 | 0 | 0 | 0 | 1 |
| 2 | $\tau^{-13}$ | 96 | 51 | 45 | 6 | 1 | 0 | 5 | 0 | 0 | 0 | 0 |
| 3 | $\tau^{-15}$ | 210 | 115 | 95 | 20 | 0 | 5 | 5 | 5 | 5 | 0 | 5 |
| 4 | $\tau^{-17}$ | 575 | 315 | 260 | 55 | 5 | 5 | 30 | 5 | 5 | 5 | 5 |
| 5 | $\tau^{-19}$ | 1,300 | 720 | 580 | 140 | 5 | 25 | 50 | 30 | 25 | 5 | 25 |
| 6 | $\tau^{-21}$ | 3,275 | 1,810 | 1,465 | 345 | 25 | 40 | 165 | 50 | 40 | 25 | 40 |
| 7 | $\tau^{-23}$ | 7,575 | 4,205 | 3,370 | 835 | 40 | 130 | 330 | 165 | 130 | 40 | 130 |
| 8 | $\tau^{-25}$ | 18,475 | 10,240 | 8,235 | 2,005 | 130 | 255 | 905 | 330 | 255 | 130 | 255 |
| 9 | $\tau^{-27}$ | 43,300 | 24,045 | 19,255 | 4,790 | 255 | 700 | 1,975 | 905 | 700 | 255 | 700 |
| 10 | $\tau^{-29}$ | 104,150 | 57,785 | 46,365 | 11,420 | 700 | 1,515 | 5,015 | 1,975 | 1,515 | 700 | 1,515 |
| 11 | $\tau^{-31}$ | 245,825 | 136,505 | 109,320 | 27,185 | 1,515 | 3,855 | 11,430 | 5,015 | 3,855 | 1,515 | 3,855 |
| 12 | $\tau^{-33}$ | 587,600 | 326,140 | 261,460 | 64,680 | 3,855 | 8,755 | 28,030 | 11,430 | 8,755 | 3,855 | 8,755 |
| 13 | $\tau^{-35}$ | 1,391,925 | 772,870 | 619,055 | 153,815 | 8,755 | 21,500 | 65,275 | 28,030 | 21,500 | 8,755 | 21,500 |
| 14 | $\tau^{-37}$ | 3,317,525 | 1,841,635 | 1,475,890 | 365,745 | 21,500 | 49,990 | 157,490 | 65,275 | 49,990 | 21,500 | 49,990 |
| 15 | $\tau^{-39}$ | 7,872,825 | 4,371,180 | 3,501,645 | 869,535 | 49,990 | 120,705 | 370,655 | 157,490 | 120,705 | 49,990 | 120,705 |

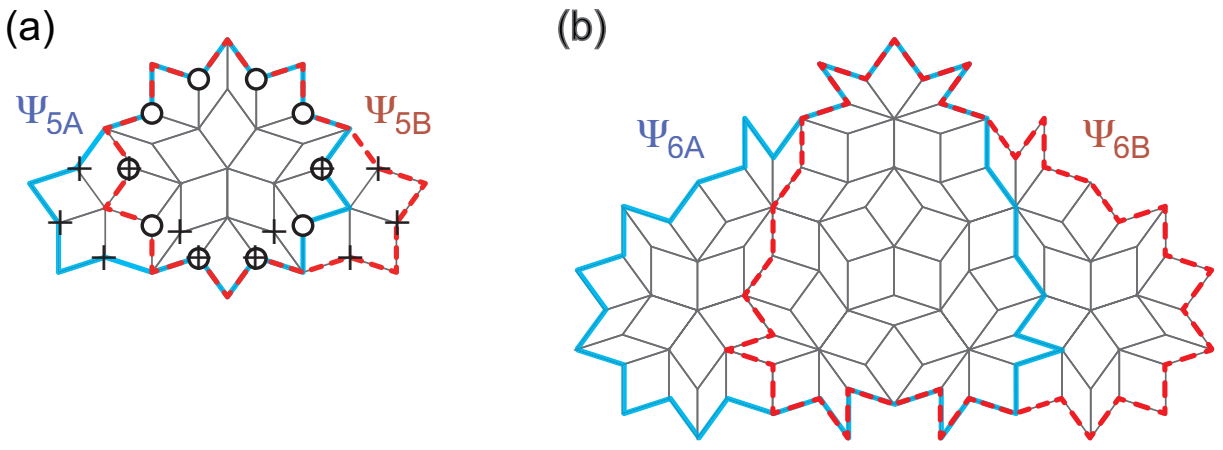

FIG. 5: (Color online) Overlap of the confined states of type-5 (a) and type-6 (b). In (a), two other confined states $\Psi_2$ and $\Psi_3$ are shown by circles and crosses, respectively.

$C_{i,A}^{(1)} = \delta_{i,1}$, $C_{i,B}^{(1)} = 0$, and $(S5)_s^{(1)} = \delta_{s,A}$.

It is clear that each cluster thus generated has the five-fold rotational symmetry inherited from the cluster-1. We can calculate the cluster densities in the infinite-size lattice, and results are represented in terms of the golden ratio $\tau$. Let us define the fraction of the cluster-$i$ by $p_i = \lim_{N\to\infty} N_{\rm cluster}^{(i)}/N$, where $N_{\rm cluster}^{(i)}$ is the number of the cluster-$i$'s in the lattice of $N$ sites. Then the recursion relation (2) leads to $p_i = \tau^{-2(i-1)} p_1$. Transformation by double deflation relates the density of S5-vertices $p_{S5}$ to the cluster-1 density as $p_1 = \tau^{-4} p_{S5}$. The vertex densities were calculated in Ref.[34] and the necessary one is $p_{S5} = 5^{-1/2}\tau^{-7}$. This results in

$$p_i = 5^{-1/2}\tau^{-2i-9}. \quad (3)$$

Next, we examine the number of the independent confined states $N_i^c$ in the cluster-$i$. We note that all confined states with the topology shown in Fig. 2 are not linearly independent. In the cluster-$i$ with $i \geq 4$, confined states of type-5 or type-6 sometimes overlap as shown in Fig. 5. In such a case, one of them can be represented as a linear combination of other confined states. For example, two confined states $\Psi_{5A}$ and $\Psi_{5B}$ of type-5 overlap in the region shown in Fig. 5 (a). There also exist two other confined states $\Psi_2$ and $\Psi_3$ of type-2 and 3, respectively. Examining their wave functions, we find the relation $\Psi_{5B} = \Psi_{5A} - 10\Psi_3 - 6\Psi_2$, and therefore these four states are not linearly independent. We also find a similar relation between the two type-6 confined states in Fig. 5 (b), and their relation is explicitly shown in Appendix B. Therefore, it is necessary to count the net number of confined states with carefully dropping redundant ones, by taking into account the topology in the lattice.

We numerically examine profiles of the clusters-$i$ up to $i = 15$ and list the statistics of sites in Table I. For the cluster-$i$, the number of M- and D-sites are denoted by $N_i^\alpha$ ($\alpha = M, D$), and the total number of sites is $N_i = N_i^M + N_i^D$. For the same clusters, we also numerically count the number of type-$k$ confined states $N_{i,k}$ for all $k$ and also list the results in the same table.

From these data, we find that these numbers follow the same recursion relation

$$X_i = 2X_{i-1} + 3X_{i-2} - 5X_{i-3}. \quad (4)$$

This holds when $i \geq 5$ for most of the numbers, but the exception is $N_{i,4}$ and $i \geq 6$ in this case. We also find that the number of the confined states is related to that of the S5-vertices in the same or smaller cluster. For example, $N_{i,1} = (S5)_{i-1}$, $N_{i,2} = (S5)_i$, and the whole list is shown in Appendix C. These relations should hold for general $i$, since all the six types of confined states have finite domains and these domains are generated from an S5-vertex in at most five deflations. Therefore, the recursion relations of $N_{i,k}$ are the consequence of the relation of $(S5)_i$. We solve the above recursion relation and the general terms of $N_i^{\alpha}$ and $N_{i,k}$ are explicitly given together with the result of $(S5)_i$ in Appendix C. One should note that the result of $\{N_{i,k}\}$ has an ambiguity, because when multiple confined states are dependent it is not unique which states should be counted. However, the sum of the results for the six types is unique and it is given as

$$\sum_{k=1}^{6} N_{i,k} = 3(S5)_i + 8(S5)_{i-1} + 5(S5)_{i-2} = \sum_{l=0}^{2} \frac{4\alpha_l^2 + 3\alpha_l - 6}{\frac{11}{5}\alpha_l^2 + 4\alpha_l - 10}\,\alpha_l^i, \quad (5)$$

where $\alpha_l$'s are roots of the characteristic equation (C2) introduced in Appendix C and their values are given in Eq. (C3).

We are ready to prove Arai's conjecture and show that all the confined states are exhausted by the six types in Fig. 2. We can also show that all confined states have zero amplitude at every D sites in any cluster, and this is proved in Appendix D. Thus, the number of independent confined states in the cluster-$i$ is given by $\Delta N_i^{AB} = N_i^M - N_i^D$. This value is calculated in Eq. (C6) and agrees with the above result of $\sum_{k=1}^{6} N_{i,k}$. This means that there exists no other confined states that are independent from the known six types. The fraction of independent confined states is thus given by

$$p_{\rm conf} \equiv \sum_{i=1}^{\infty} p_i \Delta N_i^{AB} = -50\tau + 81 \approx 9.83006 \times 10^{-2} \quad (6)$$

and this confirms the estimate in Ref. [14].

The fractions of the sites belonging to the cluster-$i$ are given by $n_i = p_i N_i$, and their cluster size dependence is plotted in Fig. 6. We find that the cluster-2 has the largest fraction $n_2 = 0.0824$ and $n_i$ decreases with cluster size as $n_i \sim (\alpha_0/\tau^2)^i$ with the decrease ratio $\alpha_0/\tau^2 \sim 0.908$. Therefore, a large system size is necessary to discuss the effects of large clusters, which may be important at weak couplings ($U/t \ll 1$). We have checked that the relation $\sum_{i=1}^{\infty} n_i = 1$ holds exactly, and this means that every site belongs to one of the clusters in the infinite-size lattice.

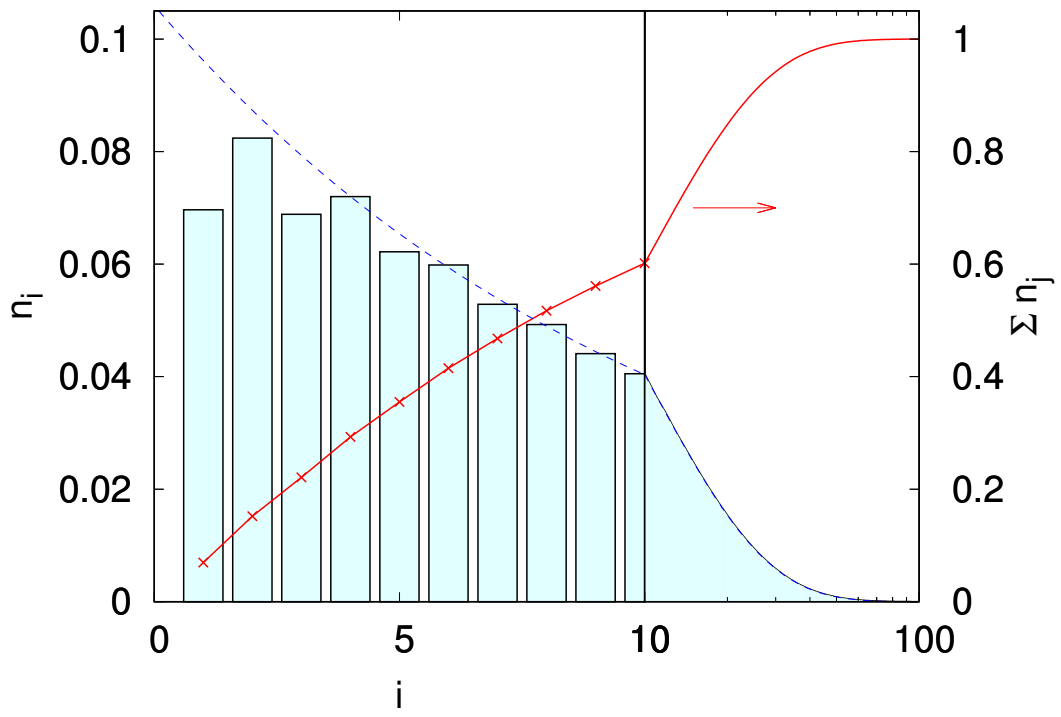

FIG. 6: The fraction of the sites in the cluster-$i$. The dashed line represents the asymptotic curve $\sim (\alpha_0/\tau^2)^i$ and the solid line with crosses represents the summed fraction $\sum_{j=1}^{i} n_j$.

## IV. ANTIFERROMAGNETIC ORDER IN THE $U \to +0$ LIMIT

Now, we study magnetic properties in the weak coupling limit. In this limit, the non-interacting electron spectrum $D(E)$ is an important factor stabilizing magnetic orders, and the Penrose-Hubbard model is special in two ways. First, confined states contribute a $\delta$-function peak at $E = 0$ in $D(E)$, and secondly this peak is separated by a finite energy gap from the continuous spectrum of extended states. The gap size is estimated $\Delta/t \approx 0.163 \pm 0.007$ in the first study [10] and our diagonalization calculation finds $\Delta/t = 0.1703$ in the system of $N = 198,771$ with the open boundary condition (OBC). We have also used the inverse iteration method to estimate the gap in larger systems. The result is $\Delta/t = 0.1685$ for the system of $N = 24,504,046$ with OBC, and this agrees well with the other estimates [35]. The total fraction of confined states is $p_{\rm conf} \approx 0.0983$, and the ground state is thermodynamically degenerate at exactly $U = 0$. Consider a finite-size lattice with $N$ sites and $N_c$ confined states. Including the spin degrees of freedom, the ground state degeneracy is $W(N_c) = (2N_c)!/(N_c!)^2$ and thus the residual entropy density is

$$\mathscr{S}_{\rm res} = \frac{1}{N}\log W(N_c) \overset{N\to\infty}{\longrightarrow} p_{\rm conf}\log 4 \quad (7)$$

Switching on an infinitesimally small Coulomb interaction lifts this degeneracy.

Since the Penrose lattice is bipartite, we expect that the ground state has an antiferromagnetic order at half filling for any $U > 0$ in the thermodynamic limit. This depends on the behavior of $D(E)$ around the Fermi energy $E = 0$. In ordinary cases with finite $D(0)$, the staggered moment of the antiferromagnetic order scales with the Coulomb interaction $U$ as $m_s \sim \exp(-{\rm const.} \times [D(0)U]^{-1})$ (see, for example, Ref. [36]). Since the Penrose-Hubbard model has a very unique $D(E)$, formation of local magnetizations takes place mainly in the part of confined states for $U \ll \Delta$, and we will show that the confined states are completely polarized in each cluster even in the $U \to +0$ limit.

Let us first examine the possibility of ferrimagnetic order. This order occurs if the A- and B-sublattices have different densities of sites, but we can exclude this possibility. As will be explained in detail later, the Penrose lattice can be generated by projecting a part of the five-dimensional lattice points onto the two-dimensional physical space $(x,y)$ [27, 28]. The projection onto the three-dimensional perpendicular space $(\tilde{x},\tilde{y},\tilde{z})$ has a one-to-one correspondence with $(x,y)$, and it contains the information on local environment of the site at $(x,y)$. In the perpendicular space, $\tilde{z}$ takes an integer value in the region $0 \le z \le 3$ and the projected image forms a regular pentagon in the $(\tilde{x},\tilde{y})$ space for each $\tilde{z}$-plane. [See also Fig. 12 (d) in Sec. VI B]. The parity of $\tilde{z}$ determines the sublattice of the site $(x,y)$: A-sublattice if $\tilde{z}$=1 or 3, and the B-sublattice if $\tilde{z}$=0 or 2. Since the projection to the $(\tilde{x},\tilde{y})$ space is uniform, the A/B-sublattice densities are proportional to $P(1)+P(3)$ and $P(0)+P(2)$ respectively, where $P(\tilde{z})$ denotes the area of the projected pentagon image in the $\tilde{z}$-plane. The projection in $\tilde{z}=2$ is the inverted image of the pentagon in $\tilde{z}=1$, and the one in $\tilde{z}=0$ is also the inversion of the pentagon in $z=3$. Therefore, $P(1)+P(3)=P(0)+P(2)$ and the A/B-sublattices have the same density in the thermodynamic limit. This proves that the antiferromagnetic order in the Penrose lattice is the ordinary type with zero uniform magnetization.

Each cluster has several confined states in M sites as discussed before, and each of the cluster-$i$'s acquires a total magnetization $\mathscr{M}_c^{(i)} = \pm\frac{1}{2}\Delta N_i^{AB}$ due to complete spin polarization of its confined states. The sign of the magnetization is determined by whether the M sites are A- or B-sublattice. As explained Sec. III, the M-sites change the A/B-sublattice from one cluster to its neighboring clusters [14]. This means that the spins in neighboring clusters are reversed. To clarify how the antiferromagnetically ordered state is realized in the Penrose lattice, we examine the fraction $p_{i\sigma}$ of the cluster-$i$ with spin $\sigma$. We can also calculate this from the recursion relation (2) with taking account of the sublattice of clusters. We choose the spin direction such that $\sigma =\uparrow$ in the A-sublattice and $\sigma =\downarrow$ in the B-sublattice: thus $p_{i,\uparrow} = p_{i,A}$ and $p_{i,\downarrow} = p_{i,B}$. Analyzing the stationary distribution of the recursion relation, we find that $p_{i,s}$ does not depend on the sublattice in the thermodynamics, and therefore

$$p_{i\sigma} = \frac{1}{2}p_i = \frac{\tau^{-2i-9}}{2\sqrt{5}}, \tag{8}$$

This result means that the equivalence between the two sublattices is realized not only in the site number density but also in the level that each cluster-$i$ appears in both sublattices with the same density. Since a net magnetization is always zero for each cluster-$i$, an antiferromagnetically ordered state is indeed realized in the Penrose Hubbard model in the thermodynamic limit.

Now, we calculate the spatial distribution of the magnetization in the Penrose Hubbard model in the weak coupling limit. Since the confined states are isolated in clusters, one can calculate local magnetizations separately in each cluster. However, the confined states shown in Fig. 2 are not orthogonal, and one need to care this point.

The density of the all confined states at the site $\mathbf{r}_j$ in the cluster-$i$ is defined as $2m_j = \sum_\alpha |\psi_\alpha(\mathbf{r}_j)|^2$ with the orthonormal wave functions of the confined states $\psi_\alpha$ and $\alpha = 1,\cdots,N_i^M - N_i^D =: N_i^c$. In principle, one may generate $\{\psi_\alpha\}$ by performing the Gram-Schmidt orthogonalization starting from those in Fig. 2 with integer values of wave function amplitude, which are denoted by $\{\tilde{\psi}_\alpha\}$. However, we can circumvent this complication and calculate $m_j$ more straightforwardly as follows. Let define a rectangular matrix $\tilde{W}$ by the nonorthogonal confined states, and its partner $W$ by the corresponding orthonormalized states

$$\tilde{W} = \begin{bmatrix} \tilde{\psi}_1(\mathbf{r}_1) & \cdots & \tilde{\psi}_{N_i^c}(\mathbf{r}_1) \\ \vdots & \ddots & \vdots \\ \tilde{\psi}_1(\mathbf{r}_{N_i^M}) & \cdots & \tilde{\psi}_{N_i^c}(\mathbf{r}_{N_i^M}) \end{bmatrix}, \tag{9}$$

$$W = \begin{bmatrix} \psi_1(\mathbf{r}_1) & \cdots & \psi_{N_i^c}(\mathbf{r}_1) \\ \vdots & \ddots & \vdots \\ \psi_1(\mathbf{r}_{N_i^M}) & \cdots & \psi_{N_i^c}(\mathbf{r}_{N_i^M}) \end{bmatrix} = \tilde{W}Q. \tag{10}$$

Here, the transformation matrix $Q$ performs the Gram-Schmidt orthogonalization and it is to be determined such that $W^\dagger W$ is the identity matrix. While $Q$ is a square matrix with size $N_i^c \times N_i^c$, the size of $\tilde{W}$ and $W$ is $N_i^M \times N_i^c$ and $N_i^M > N_i^c$. The overlap matrix $\tilde{W}^\dagger\tilde{W}$ is real symmetric and positive definite, and therefore can be diagonalized with an orthogonal matrix $P$

$$\tilde{W}^\dagger\tilde{W} = P\,\mathrm{diag}\left(\Delta_1,\cdots,\Delta_{N_i^c}\right){}^tP, \tag{11}$$

where all the eigenvalues $\{\Delta_k\}$ are strictly positive, and ${}^tP$ is the transpose of the matrix $P$. Then, we can choose the transformation matrix as follows

$$Q = P\,\mathrm{diag}\left(\Delta_1^{-1/2},\cdots,\Delta_{N_i^c}^{-1/2}\right), \tag{12}$$

and this orthonormalizes the confined states. It is straightforward to calculate the confined state density with these matrices

$$\begin{aligned} 2m_j &= \sum_\alpha |\psi_\alpha(\mathbf{r}_j)|^2 = \left(WW^\dagger\right)_{jj} = \left(\tilde{W}QQ^\dagger\tilde{W}^\dagger\right)_{jj} \\ &= \left[\tilde{W}(\tilde{W}^\dagger\tilde{W})^{-1}\tilde{W}^\dagger\right]_{jj}. \end{aligned} \tag{13}$$

Therefore, it is not necessary to obtain the transformation matrix $Q$ and one can calculate $m_j$ directly from the nonorthogonal wave functions of the confined states.

This expression also immediately shows that all the $m_j$ values are rational numbers. Recall that $\tilde{W}$'s matrix elements are all integer. Therefore the elements of $(\tilde{W}^\dagger\tilde{W})^{-1}$ are all rational numbers, and proves that $m_j$ is rational. For examples, the magnetization of certain sites in the cluster-1 is $1/20$, and those in cluster-2 are $227/4920, 2/41$, and $331/2460$. The distributions of magnetizations in the clusters-1, 2, and 3 are shown in Figs. 7 (a), (b) and (c), respectively. The five-fold rotational symmetry appears in the magnetization distribution. By performing similar calculations, we obtain the distribution of the magnetization in Fig. 7 (d). Due to the numerical restriction, we have calculated the magnetizations up to the cluster-12 with 587,600 sites, implying that seventy percents of the magnetization are exactly obtained. Namely, the average of the staggered moment is given by the number of confined states as

$$m_0 = \frac{1}{2N}\sum_{i\sigma}(N_i^M - N_i^D)p_{i\sigma}$$

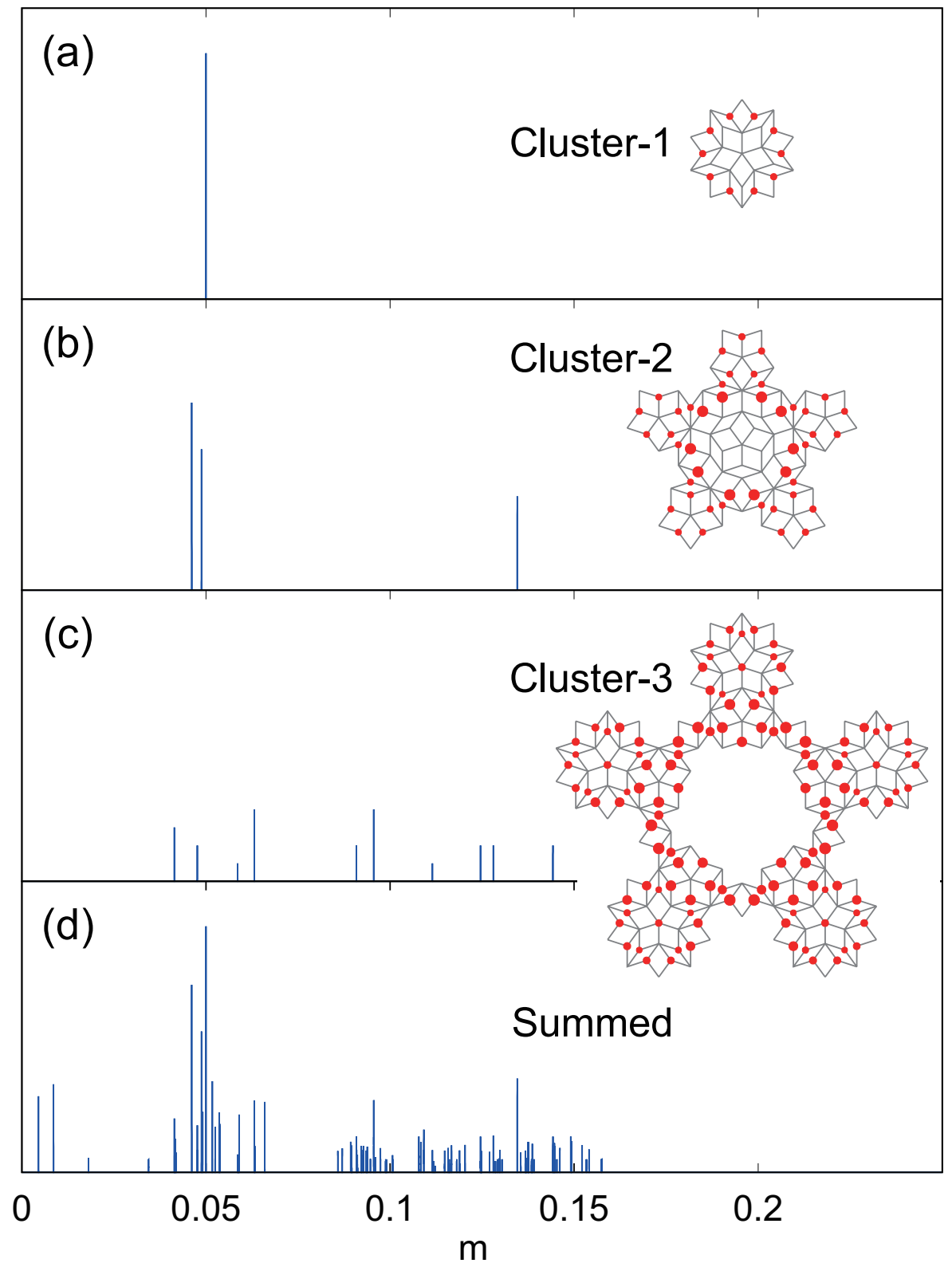

FIG. 7: (Color online) Local magnetizations in the Penrose-Hubbard model at $U = +0$. (a), (b), and (c) show the magnetization for the cluster-1, 2, and 3. Spatial pattern for the magnetization is shown in the insets. (d) shows the distribution of the magnetizations summed up to the cluster-12.

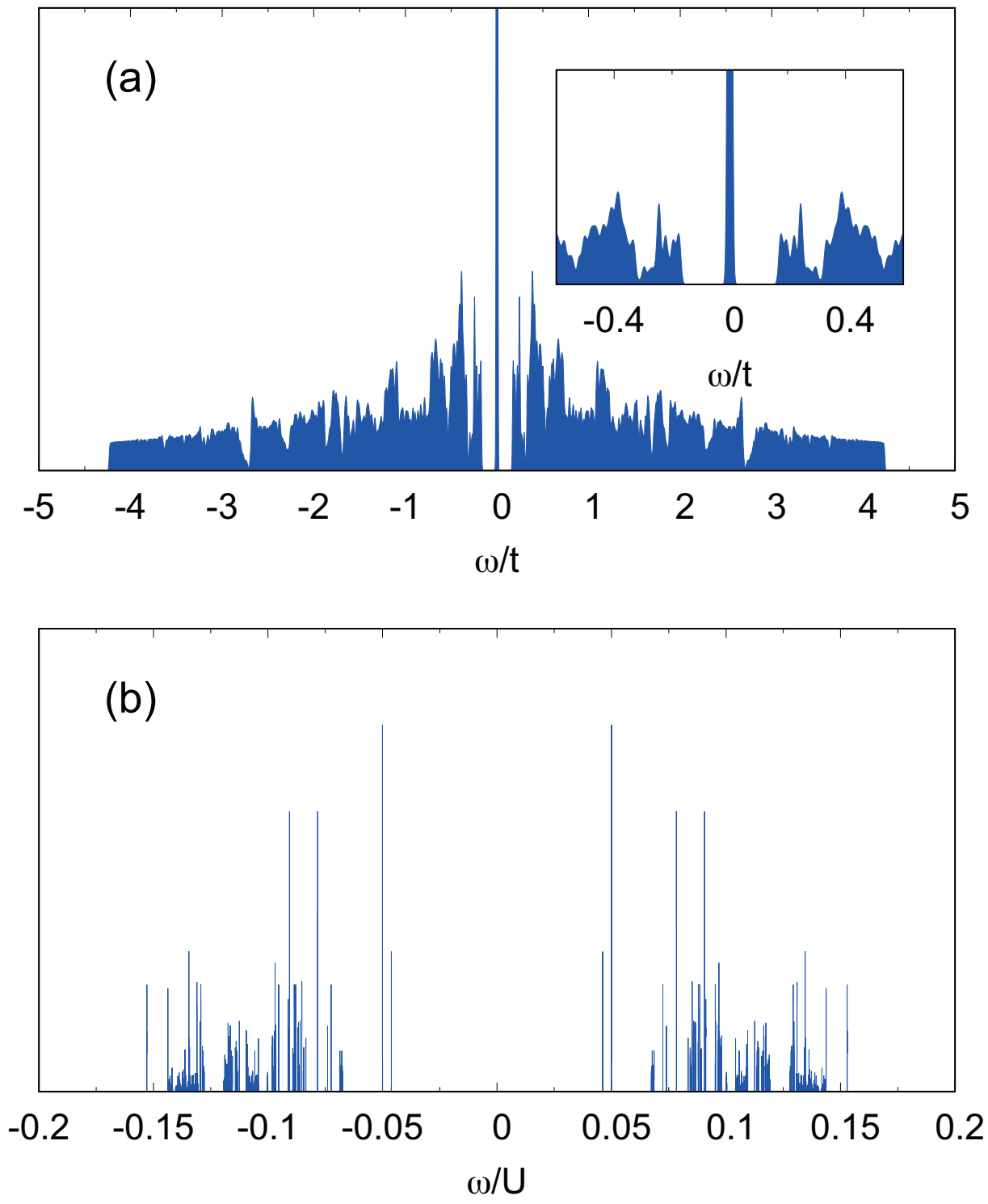

FIG. 8: (Color online) (a) DOS in the Hubbard model in the weak coupling limit on the Penrose lattice with $N = 198,771$ sites. Inset is a magnified figure of the central part. (b) DOS of the confined states in the weak coupling limit. Their energies are proportional to $U$, and thus the renormalized values are used. The results up to the cluster-10 are shown.

$$= \; -25\tau + \frac{81}{2} \sim 0.04915. \tag{14}$$

As discussed above, switching the Coulomb interaction yields the finite magnetization. At $U = +0$, this lifts the macroscopic degeneracy in the ground state and the staggered magnetization jumps rather than continuous evolution. The delta-function peak at $E = 0$ in the DOS should split into multiple peaks, but this split is continuous and proportional to $U$ in the weak coupling limit. We note that, when the antiferromagnetically ordered state is realized, the effective potential for electrons is given via the magnetic distribution. Therefore, we again consider the clusters independently to calculate the energy shift for the confined states. The effective potential is proportional to the interaction strength as

$$V_{\alpha\alpha'} = U \sum_j \psi_\alpha(\mathbf{r}_j) m_j \psi_{\alpha'}(\mathbf{r}_j). \tag{15}$$

Performing the first-order perturbation theory in the potential, we obtain the energy shifts for confined states in each cluster. The cluster-1 has one confined state and only one value of magnetization $m_j = \pm 1/20$. This shifts the confined-state energy to $E/U = \pm 1/20$, where the sign depends on if the cluster has the M sites in the A or B sublattice. The cluster-2 has six confined states, and their energies split to four values $E/U = \pm 227/4920, \pm(8521 \pm 277\sqrt{5})/100860, \pm 331/2460$. The reason of the four values is as follows. Out of the six confined states, one is type-1 and the other five are type-3. The latters are labeled by the angular momentum about the center $L_z$=0, $\pm 1$, and $\pm 2$, because of the five-fold symmetry of the cluster. The local magnetizations keeps this symmetry, and thus the wave functions do not change, except in the $L_z = 0$ sector where the symmetric type-3 state hybridizes the type-1 state. Because the perturbation $V_{\alpha\alpha'}$ is real, the energies of the $\pm L_z$ states are degenerate. Therefore, among the four values, the two energies including $\sqrt{5}$ are doubly degenerate. The DOS of the confined states are shown in Fig. 8. We have clarified that the introduction of the Coulomb interaction divides the single peak at $E = 0$ into many peaks in the DOS, and their distribution is reflected by the confined state inherent in the Penrose lattice.

In this section, we have defined the clusters in the Penrose lattice, which is based on the confined states discussed in Ref. [14]. Counting the number of confined states and sites in each cluster, we have confirmed that the kinds of the confined states are restricted to be six (the Arai's conjecture). Furthermore, we have clarified that the numbers of the sublattices A and B are the same, and the antiferromagnetically ordered ground state is realized without the uniform magnetization in the thermodynamic limit. We have also performed numerical calculations to obtain the distribution of the magnetization

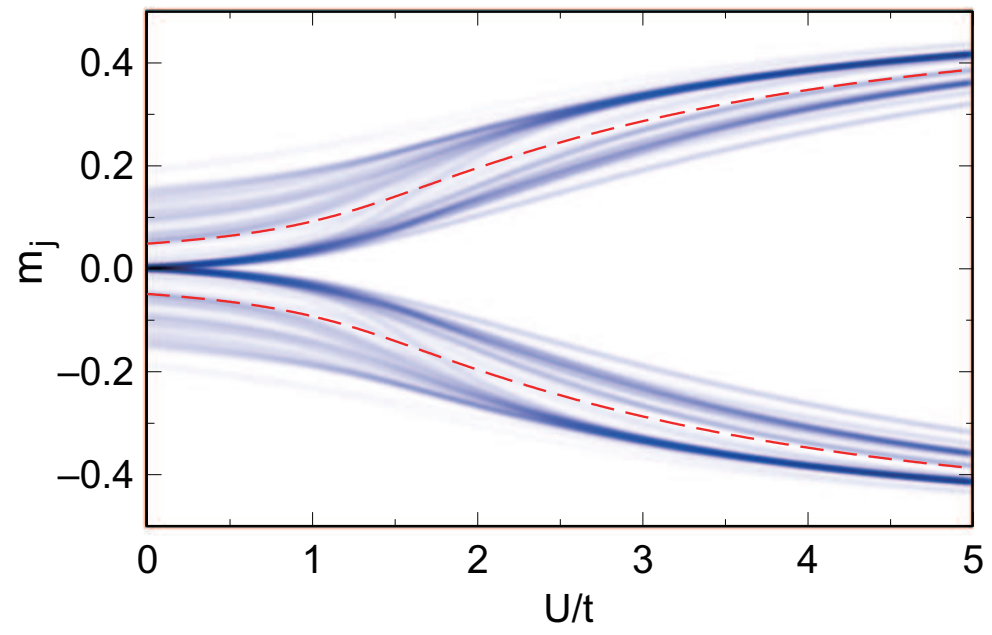

FIG. 9: Distribution of local magnetizations as a function of the Coulomb interaction $U/t$ in the system with $N = 75,806$. The sublattice averages $\bar{m}_A$ and $\bar{m}_B$ are also shown by dashed lines.

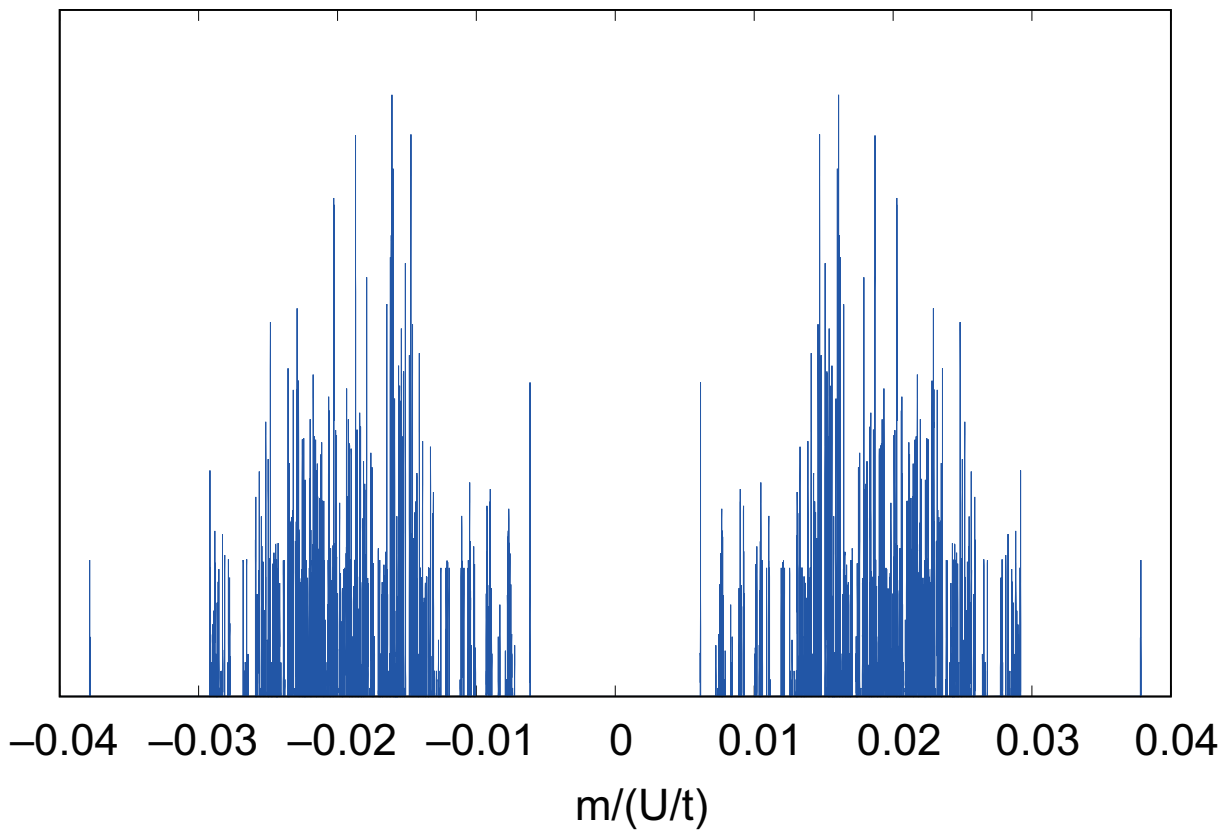

FIG. 10: Distribution of the induced magnetizations normalized by $U/t$ in the system with $N = 198,771$.

and DOS in the weak coupling limit.

## V. EFFECTS OF FINITE $U$

In the section, we consider finite $U$ cases and study how the antiferromagnetic order evolves in the ground state with increasing $U$ from the weak coupling limit discussed in the previous section. Here, we make use of the unrestricted Hartree-Fock approximation and the Hamiltonian (1) is transformed to

$$H_{\rm HF} = -t \sum_{\langle j,j'\rangle\sigma} c^\dagger_{j\sigma} c_{j'\sigma} - \sum_j h_j S^z_j, \qquad (16)$$

$$S^z_j \equiv \tfrac{1}{2}(n_{j\uparrow} - n_{j\downarrow}),\ h_j = 2Um_j \equiv 2U\langle S^z_j\rangle, \qquad (17)$$

where the local magnetizations $m_j \equiv \langle S^z_j\rangle$ are to be determined self-consistently at all the sites. In our calculation, we use the open boundary condition, and examine finite lattices with $N = 75,806$ and $198,771$, where the largest clusters are the cluster-8 and cluster-9, respectively. These lattices are generated by repeating deflation from the star structure with an S5-vertex center shown in the right side of Fig. 1 and thus have the global five-fold rotational symmetry. For given values of $\{m_j\}$, we numerically diagonalize the Hamiltonian $H_{\rm HF}$ and update the magnetizations, and repeat this procedure until the result converges.

Due to the quasiperiodicity of the lattice, the local magnetization is not uniform and differs from site to site. The magnetization distribution changes with $U$ and Fig. 9 shows its behavior. Here, the symmetrized distribution is shown since the finite system does not have the symmetry in distribution for $m \to -m$. For $U = +0$, the distribution agrees with the analysis in the previous section except for minor edge effects: many sites have $m_j = 0$, while the average staggered magnetization is finite. $\bar{m}_A = N_A^{-1}\sum_{j\in A} m_j$ is the average moment in the A-sublattice and $\bar{m}_B$ is the B-sublattice average. Increasing Coulomb interaction induces finite magnetizations at the sites previously nonmagnetic. Its $U$-dependence drastically differs from that in the conventional case of bipartite lattices, where the staggered magnetization usually increases as $m_s \sim \exp(-\text{const.} \times [D(0)U]^{-1})$ with $D(0)$ being the DOS at the Fermi energy. In the Penrose-Hubbard model, the mean field in Eq. (16) is finite and asymptotically $h_j \propto U$ on many M sites, because the confined states have a finite spin polarization there already at $U = +0$. This mean field polarizes spins of extended states, and the induced moment at the previously nonmagnetic sites has a size proportional to the field strength, $m_j \sim \sum_{j'} \chi_{jj'} h_{j'}$, and therefore $m_j \propto U/t$. In the weak coupling limit, $h_{j'}/U$ is finite only at those M sites where the confined states have a finite amplitude. $\chi_{jj'}$ is the nonlocal susceptibility of the extended electrons between the sites $j$ and $j'$: $\chi_{jj'} \geq 0$ if the two sites are in the same sublattice and $\chi_{jj'} \leq 0$ otherwise. This is proved in Appendix E. One should note that the main part of the mean field is the staggered component, but many other spatial components are also mixed, because of the distribution of magnetizations at $U = +0$. Thus, the induced moment is a superposition of the responses of these spatial components, and exhibits a complicated pattern.

Figure 10 shows the distribution of $m_j/(U/t)$ at those sites nonmagnetic at $U = +0$. The local magnetizations are calculated for the small value $U = 10^{-5}t$ in the lattice with $N = 198,771$. Due to its finite-size effects, the total magnetization does not vanish $\sum_j m_j \neq 0$, and therefore we show the symmetrized distribution in the figure. Comparison with the similar plot for another $U$ value confirms the scaling $m_j \propto U/t$ of the induced magnetizations. It is found that the magnetic moments distribute mainly in the region $0.01 \leq |m_j|/(U/t) \leq 0.03$ and the average is $|m_j|/(U/t) = 0.0189$.

In the strong coupling region, the system is reduced to the Heisenberg model on the Penrose lattice with the nearest-neighbor exchange coupling $J = 4t^2/U$. The mean-field ground state has a uniform staggered moment $m_j = \pm\frac{1}{2}$ at $U = \infty$. This differs from the result of the spin wave theory for the Heisenberg model that predicts site-dependent reduction of the moment. This reduction is due to inhomogeneous quantum fluctuations that are mainly determined by the coordination number of each site [21, 22]. This effect is not taken into account in the mean-field approximation for the Hubbard model, and one needs a more elaborate approach like the random-phase approximation, but this is beyond the scope of the present study. Despite this limitation, it is interesting

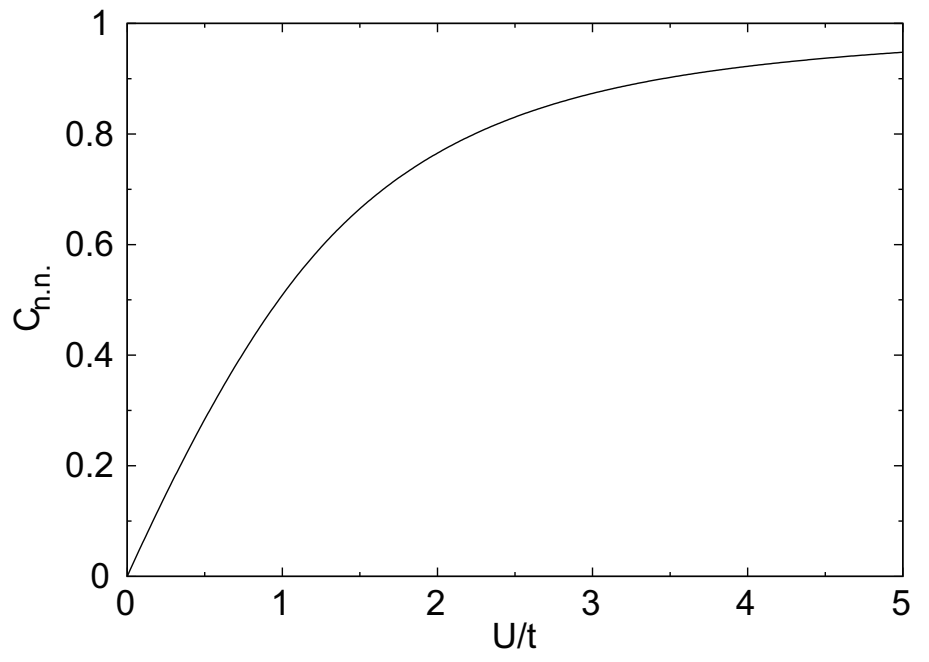

FIG. 11: Nearest-neighbor spin correlation.

that the present approach demonstrates a common feature as will be shown later.

The antiferromagnetic order evolves continuously upon increasing $U$, but their spatial profile differs between the weak and strong coupling limits. One characteristic quantity is the nearest-neighbor spin correlation

$$C_{\text{n.n.}} \equiv \frac{1}{N_b} \sum_{\langle j,j' \rangle} \frac{m_j m_{j'}}{\bar{m}_A \bar{m}_B}. \tag{18}$$

Here, the sum is taken over all the nearest-neighbor site pairs and it is normalized by this pair number $N_b$. Figure 11 plots the variation of $C_{\text{n.n.}}$ with the interaction strength. We find that the nearest-neighbor spin correlation starts from zero. This fact contrasts with the behavior of the Hubbard model on simple lattices such as the square or cubic lattice. The staggered magnetization is uniform in those lattices and $C_{\text{n.n}} = 1$ in the mean field approximation. Therefore, the linear dependence of the nearest-neighbor spin correlation originates from two effects. One is the finite value of magnetization in many M-sites in clusters already at $U = +0$. The other is the fact that at their nearest-neighbor D-sites magnetization grows linearly with $U$ in the weak coupling region. For $U \gg t$, $C_{\text{n.n.}}$ approaches unity since the staggered magnetization becomes asymptotically uniform in the lattice.

## VI. MAGNETIZATION PROFILE

### A. Perpendicular space

Now, let us substantiate a quasiperiodic feature in the spatial profile of local magnetizations. To this end, it is useful to analyze the profile in the perpendicular space [37]: positions in the perpendicular space have one-to-one correspondence with $\{\mathbf{r}_j\}$'s in the physical space. Sites in the Penrose lattice correspond to a subset of the five-dimensional lattice points $\vec{n} = (n_1, n_2, n_3, n_4, n_5)$ labeled with integers $n_\mu$ and their coordinates are the projections onto the two-dimensional physical space

$$\mathbf{r} = (x,y) = (\vec{n}\cdot\vec{e}^{\,x}, \vec{n}\cdot\vec{e}^{\,y}) = \sum_{\mu=1}^{5} n_\mu \mathbf{e}_\mu, \tag{19}$$

where $e^x_\mu = \cos(\phi\mu + \theta_0)$, $e^y_\mu = \sin(\phi\mu + \theta_0)$ with $\phi = 2\pi/5$, and $\mathbf{e}_\mu = (e^x_\mu, e^y_\mu)$. The initial phase $\theta_0$ is arbitrary, and an example with the choice $\theta_0 = \pi/10$ is shown in Fig. 1, where $\{\mathbf{e}_\mu\}$'s are also depicted. The projection onto the three-dimensional perpendicular space has information specifying the local environment of each site [27],

$$\tilde{\mathbf{r}} = (\tilde{x}, \tilde{y}) = (\vec{n}\cdot\tilde{\vec{e}}^{\,x}, \vec{n}\cdot\tilde{\vec{e}}^{\,y}), \;\; \tilde{z} = \vec{n}\cdot\tilde{\vec{e}}^{\,z}, \tag{20}$$

where $\tilde{e}^x_\mu = \cos(2\phi\mu)$, $\tilde{e}^y_\mu = \sin(2\phi\mu)$, and $\tilde{e}^z_\mu = 1$. $\tilde{z}$ takes only four values $\{0,1,2,3\}$, and in each $\tilde{z}$-plane the $\tilde{\mathbf{r}}$-points densely cover a region of pentagon shape. Moreover, the pentagon in $(3-\tilde{z})$-plane has the same size as the one in $\tilde{z}$-plane, but it is reversed upside down. Note that the sites with odd or even number $\tilde{z}$ correspond to the A/B sublattices, since upon moving from one site to its neighbor only one of $n_\mu$'s changes by $\pm 1$. Therefore, the antiferromagnetically ordered state may also be characterized by an alternate sign of magnetization in the four $\tilde{z}$-planes; essentially $m(\tilde{\mathbf{r}}, \tilde{z}) = -m(-\tilde{\mathbf{r}}, 3-\tilde{z})$, but one should note that $\tilde{\mathbf{r}}_j$ positions do not match exactly between the two planes if the lattice size is finite.

### B. Magnetization profile in the perpendicular space

Magnetization profile is shown in Fig. 12 for the two projected planes of $\tilde{z} = 1$ and 3. They are both in the A-sublattice, and $m_j \geq 0$. As discussed in Sec. IV for the weak coupling limit, the existence of the confined states yields the cluster structure, where the local magnetization appears only in its M sites. In this perpendicular-space representation, sites with similar environment map into the same subdomain in the $\tilde{\mathbf{r}}$-space. In fact at $U = +0$, the magnetized sites in the cluster-1 ($m = 1/20$) are mapped into the inside of the small yellow pentagon depicted in Fig. 12(a) or one of its nine equivalent regions. Since most of the sites are disordered, one finds large black regions in Fig. 12(a), and they corresponds to the parts of J-, K-, S3-, and S4-vertices as well as most of S- and S5-vertices. When the Coulomb interaction increases, all the sites are magnetized and the profile exhibits many star and pentagon patterns with several sizes. The profile has a fractal character and each pattern also shows a subpattern made of smaller stars and pentagons in a self similar way. The self similarity is the most prominent character of quasicrystalline structure, and many physical properties also exhibit related behaviors [37–39].

Let us analyze the crossover upon increasing $U$ in more detail. In the preceding related studies on strongly correlated systems [21–23, 25, 26], the effects of the quasiperiodic lat-tice structure have been mainly discussed from the viewpoint of the distribution of the number of nearest-neighbor sites, i.e., the coordination number. In the Penrose lattice, the coordination number ranges from 3 to 7: 3 for the vertices D and Q, 4 for K, 5 for J, S and S5, 6 for S4, and 7 for S3, as shown in Fig. 1. A small coordination number leads to a narrower width in the electron local DOS and the Coulomb repulsion has larger effects. Therefore, upon increasing $U$, we expect that the local magnetization grows faster at the sites with small coordination number. On the other hand, starting from the strong coupling limit, a small coordination number enhances quantum fluctuations, and this reduces the magnetization size.

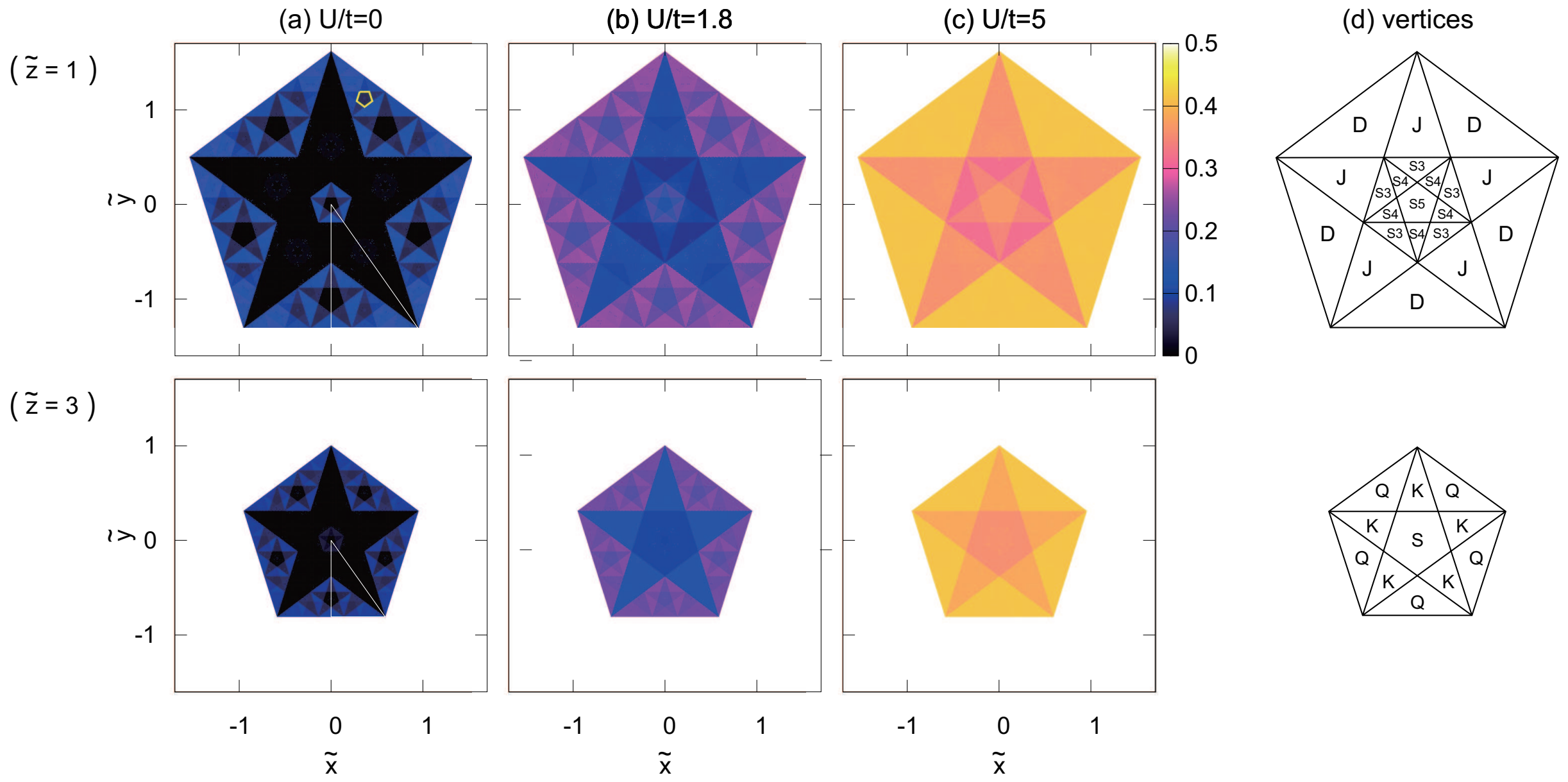

FIG. 12: Magnetization profile in the perpendicular space $(\tilde{x},\tilde{y})$ for the system with $N = 198,771$. Upper panels are in the $\tilde{z} = 1$ plane, while lower ones in $\tilde{z} = 3$. The profile has $C_{5v}$ symmetry in each pentagon region, and its principle part is the right triangle marked in (a) by white edges. $U/t =$ (a) $10^{-5}$ (essentially the same as 0) , (b) 1.8, and (c) 5.0. (d) Each part is the region of one of the eight types of vertices shown in Fig. 1.

At each site $j$, we have calculate the $U$-dependence of the local magnetization $m_j(U)$, and checked that the dependence is monotonic for $0 < U < \infty$ as expected. We then define the crossover strength $U^X_j$ by the value where the change is largest, $m''_j(U^X_j) = 0$, i.e., the inflection point of the curve. We find that $m_j(U)$ curve has only one inflection point at every site, and this uniquely defines the crossover strength $U^X_j$. Figure 13(a) shows its distribution, and the crossover strength ranges roughly from $U^X_j/t = 1.3$ to 2.4. The mean value is $\bar{U}^X/t \equiv N^{-1}\sum_j U^X_j/t = 1.75$, which is slightly larger than the crossover value $U/t \sim 1.2758$ in the square lattice model.

The crossover strength also shows a complex spatial profile. We have checked the correlation of its value and the local magnetization, $m_j(U = 0)$ or $\lim_{U\to 0} m_j(U)/U$ depending on the site, but found no clear correlation. Plotting $U^X_j$ in the perpendicular space, we now find a systematic pattern of its distribution as shown in Fig. 13(b). The magnetization value is primarily determined by the eight vertex types. The crossover strength is small at D- and Q-vertices. The sites with large $U^X_j$ are most of S- and S3-vertices in addition to some of J- and S5-vertices. However, each vertex region has a complicated pattern with self-similar character once again. There-fore, the crossover strength is determined by the local environment including not only nearest neighbors but also further distant neighbors.

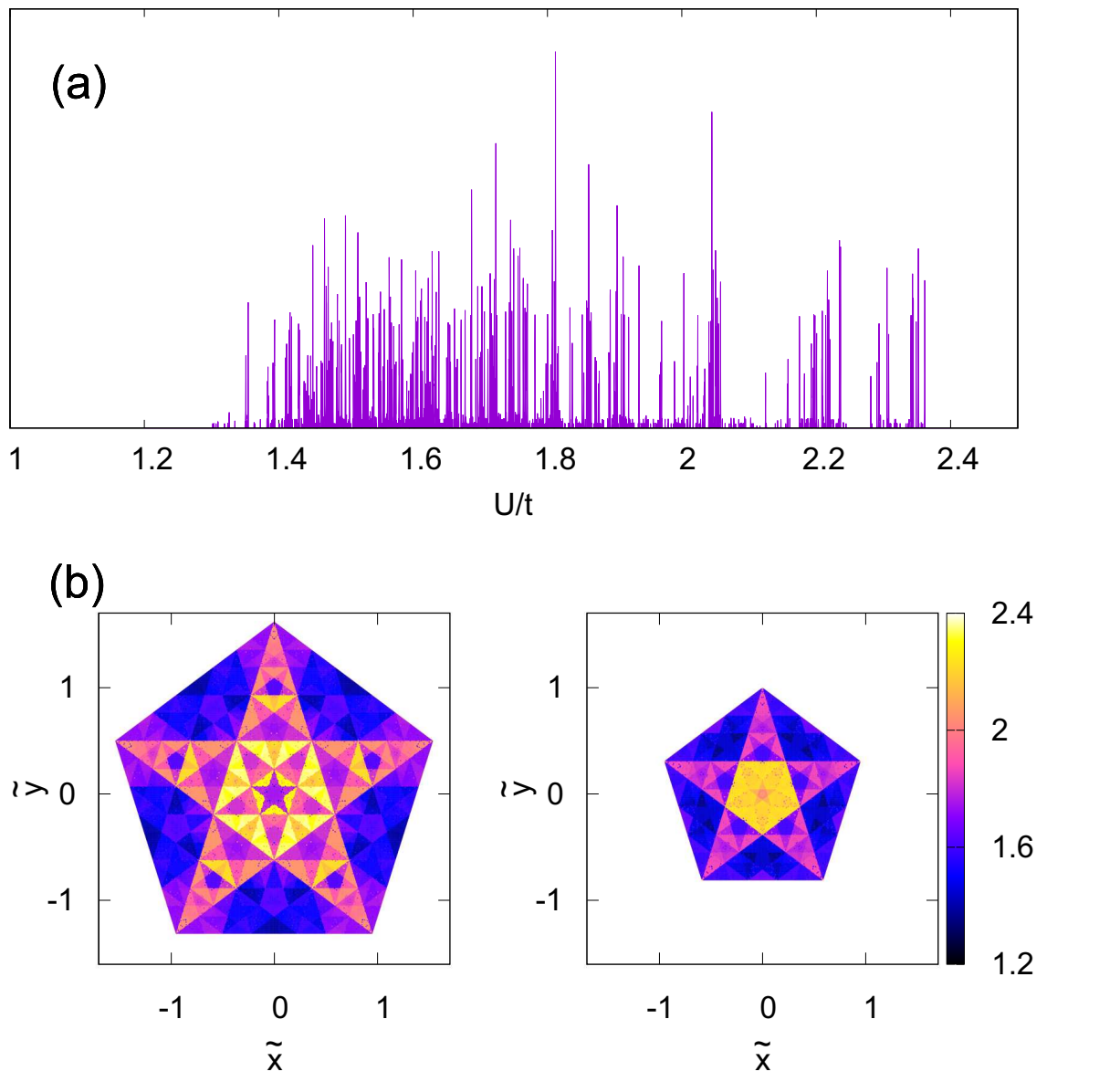

FIG. 13: (a) Distribution of the crossover strength $U^X_j$. (b) Perpendicular-space plot of $U^X_j$. Left (right) panel is for the $\tilde{z} = 1$ (3) plane.

### C. Mode analysis in the perpendicular space

Now, let us go back to the magnetization profile in Fig. 12. With varying $U$, the fractal pattern in $m(\tilde{x},\tilde{y})$ continuously changes from the weak to the strong coupling limit, and let us quantify this change by a mode analysis. Usually, Fourier transformation is a standard tool of mode analysis. The present patterns displays clear borders with geometrical shapes, particularly stars and pentagons of several sizes. Fourier transformation needs many modes to reproduce this character and hence not effective. Alternatively, we try an analysis in which the form of spatial modes is not presumed but to be automatically determined during calculation. To this end, we will use the singular value decomposition [40], which has recently become popular for its application in image compression technique.

Consider real-valued data in a two-dimensional rectangular region, and the present case is $m(\tilde{x},\tilde{y})$ in one $\tilde{z}$ plane. Discretizing the perpendicular space to grid points with number $N_{\tilde{x}} \times N_{\tilde{y}}$, the data are represented by a rectangular matrix of this size. The singular value decomposition transforms this matrix into the following form

$$m = U_{\tilde{x}} \Lambda {}^t U_{\tilde{y}}, \quad \Lambda_{\alpha\alpha'} = \delta_{\alpha\alpha'}\lambda_\alpha, \tag{21}$$

where it is assumed that $N_{\tilde{x}} \le N_{\tilde{y}}$, and $\Lambda$ is a rectangular matrix with the same size as $m$. Its diagonal elements are singular values $\{\lambda_\alpha\}$, a set of non-negative real numbers stored in the descending order $\lambda_1 \ge \lambda_2 \ge \cdots \ge \lambda_{N_{\tilde{x}}}$. $U_{\tilde{x}}$ and $U_{\tilde{y}}$ are orthogonal matrices with dimension $N_{\tilde{x}}$ and $N_{\tilde{y}}$, respectively. Their column vectors, $\{\vec{u}_H^{(\alpha)}\}$ and $\{\vec{u}_V^{(\alpha)}\}$ represent coordinate dependence of spatial modes. Note that the singular values are related to the integrated squared amplitude as $\sum_{\tilde{x},\tilde{y}} m(\tilde{x},\tilde{y})^2 = \sum_\alpha \lambda_\alpha^2$. The magnetization profile is predominantly determined by the modes with large singular value $\lambda_\alpha$

$$m(\tilde{x},\tilde{y}) \sim \sum_{\alpha=1}^{\text{cutoff}} \lambda_\alpha u_H^{(\alpha)}(\tilde{x}) u_V^{(\alpha)}(\tilde{y}). \tag{22}$$

The singular value decomposition optimizes the form of modes used in this expansion. One does not provide input of their form, and the optimization is achieved automatically. One crucial constraint is that the region should have a rectangular shape, or at least quadrangle, but does not match the pentagon region in each $\tilde{z}$ plane. One should notice that the magnetization profile has a five-fold symmetry augmented by five mirrors in the perpendicular space, inherited from the symmetry in the physical space. Therefore, instead of the entire pentagon, its one-tenth primary part is sufficient to characterize the profile. The primary part is a right triangle marked by white edges in Fig. 12(a), i.e. a half rectangle, and we fill the other half by symmetrizing data, $m(\tilde{x},\tilde{y}) = m(\tilde{x}_{\max} - x, \tilde{y}_{\min} - y)$. Here, $(\tilde{x}_{\max}, \tilde{y}_{\min})$ is the position of the bottom right corner of the primary part. We discretize the square region to the grid points with $(N_{\tilde{x}}, N_{\tilde{y}}) = (290, 400)$, and therefore the singular value decomposition determines 290 modes. With this rectangular data, we perform the singular value decomposition. By comparing the results for two systems with $N = 75,806$ and $198,771$, we have confirmed that main results shown below do not depends on the system size. We have also checked that the grid size of the magnetic pattern affects only small singular values.

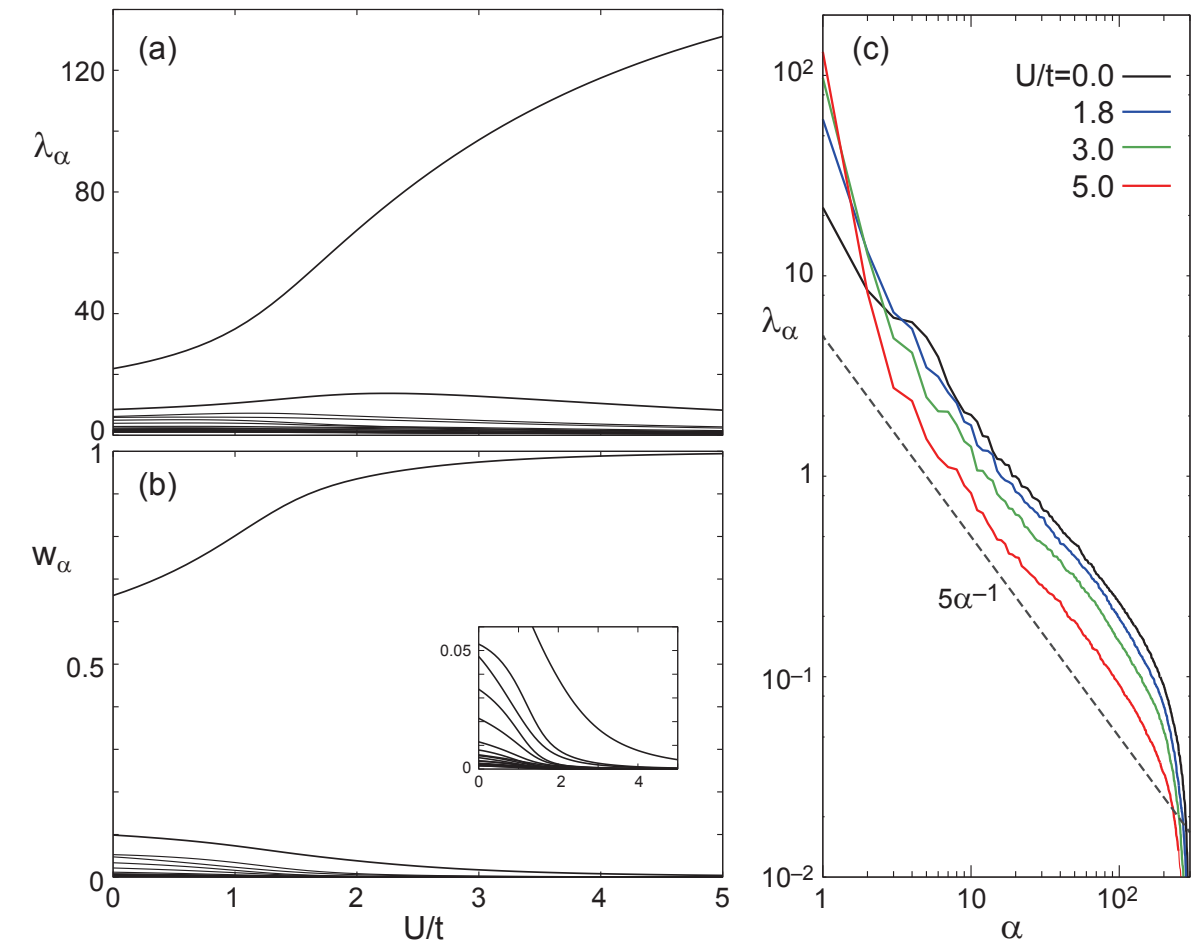

FIG. 14: (a) Twenty largest singular values of magnetization profile in the perpendicular space. (b) Their weights. Inset is a magnified graph for the small value region. (c) Size distribution of singular values.

Figure 14(a) shows the variation of 20 largest singular values of $m(\tilde{x},\tilde{y})$ in the $\tilde{z} = 1$ plane when $U$ increases. The data are calculated at each $U$ for the system with $N = 75,806$. While the largest one $\lambda_1$ predominates, $\lambda_2$ is still sizable but its variation is not monotonic. The weights of the singular values are defined as $w_\alpha = \lambda_\alpha^2 / \sum_{\alpha'} \lambda_{\alpha'}^2$. This satisfies the sum rule $\sum_\alpha w_\alpha = 1$, and its variation is shown in (b) in the same figure. At $U = +0$, the $\lambda_1$ mode already has weight $w_1 = 0.661$, while $\lambda_2$ is limited to $w_2 = 0.099$ and its variation is now monotonic. The remaining is filled by many modes. The size of $\lambda$'s is plotted in Fig. 14(c). Except a few largest and very small ones, they roughly follows the scaling $\lambda_\alpha \propto \alpha^{-1}$. This power-law behavior is consistent with the fractal profile made of self-similar stars and pentagons in Fig. 12(a). As $U$ increases, the $\lambda_1$ mode becomes even more dominant, and its weight is $w_1 = 0.920$ at $U/t = 1.8$ and 0.994 at $U/t = 5.0$, while $w_2$=0.044 and 0.004 at these $U$'s. The size distribution shows that except for a few largest ones the power-law scaling always holds for all $U$'s. The detailed analysis for some modes with larger singular values is shown in Appendix F.

### D. Spin structure factor

Before summarizing this section, we also discuss how the spatial distribution in the magnetization affects the diffraction

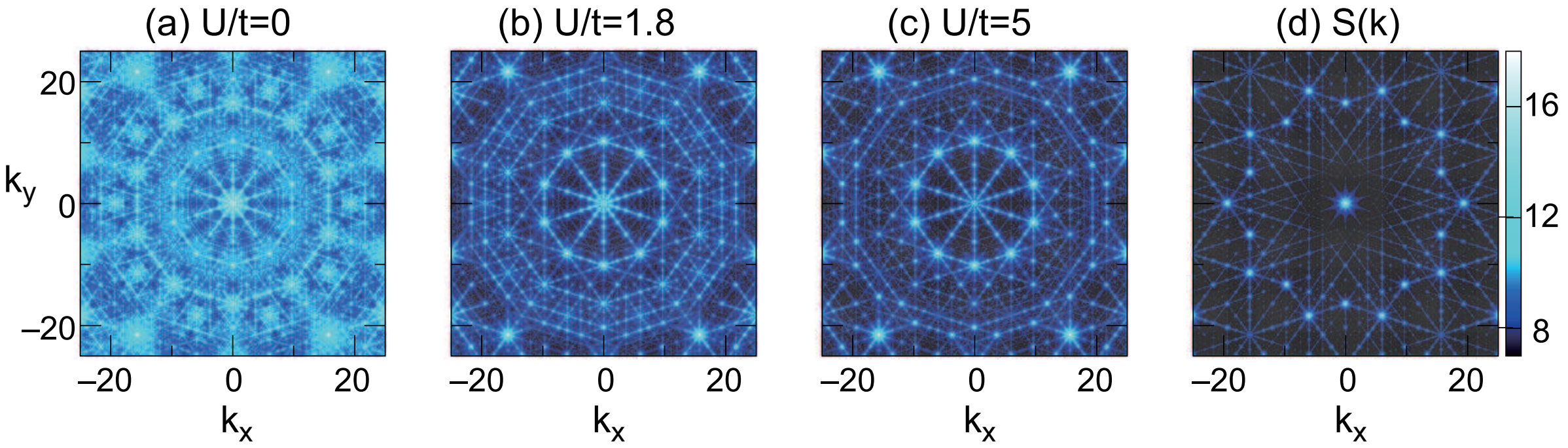

FIG. 15: (Color online) Logarithmic plot of the spin structure factor $\log[S_M(\mathbf{k})/\overline{m}^2]$ in the system with $N = 198,771$. $U/t$= (a) $10^{-5}$ (essentially the same as 0), (b) 1.8, and (c) 5. (d) is the lattice structure factor.

pattern. To this end, we calculate the spin structure factor

$$S_M(\mathbf{k}) = \left| \frac{1}{N} \sum_j m_j e^{i\mathbf{k}\cdot\mathbf{r}_j} \right|^2 = \frac{1}{N^2} \sum_{j,j'} m_j m_{j'} e^{i\mathbf{k}\cdot(\mathbf{r}_j - \mathbf{r}_{j'})}. \quad (23)$$

Because of the quasiperiodicity, the wave vector is not conserved, and we calculate the diagonal part of the structure factor. The results for $U/t$= 0, 1.8 and 5 are normalized by the mean value squared $\overline{m}^2$ and shown in Fig. 15. The lattice structure factor, $S(\mathbf{k}) = |N^{-1}\sum_j e^{i\mathbf{k}\cdot\mathbf{r}_j}|^2$ is also shown for comparison.

In the weak coupling limit $U/t \to 0$, some peaks appear due to the confined states magnetized. Increasing the interactions, the staggered magnetizations increase and some peaks develop in the **k**-space. Since all lattice sites are magnetized, the detailed structure appears in the quantity $m_{\mathbf{k}}$. At $U/t = 5$, the ten spots in $S_M(\mathbf{k})$ at $|\mathbf{k}| \sim 6$ are much stronger than those for smaller $U$. These **k**'s corresponds to twice the nearest-neighbor bonds, $\mathbf{k}_\mu \cdot (2\mathbf{e}_\mu) = \pm 2\pi$ and their large amplitudes indicate strong enhancement of antiferromagnetic spin correlation between nearest neighbor sites in the strong coupling region.

## VII. SUMMARY

In this paper, we aimed to investigate the effects of quasiperiodic lattice structure on antiferromagnetic order, and have performed a mean-field analysis for the ground state of the half-filled Hubbard model on the Penrose lattice. A crucial feature of the Penrose lattice is the presence of strictly localized states with thermodynamic degeneracy at zero energy, which have been known as confined states. These confined states lead to several exotic properties to the antiferromagnetic order. First, they have full magnetic moments at even $U = +0$, and secondly, a more surprising point is their exotic profile in space. In contrast to conventional antiferromagnetic orders, the spontaneous magnetization does not oscillate from one site to its neighboring site. Instead, the whole Penrose lattice divides up into cluster regions, and the magnetizations have the same sign within each cluster, while sites with zero confined-state amplitude have no magnetization. These clusters have a power-law distribution in their size and density: the size starts from 31 sites for the smallest one, then 96 for the next, and continues to increase up to infinity. The cluster density has the highest value $\tau^{-11}/\sqrt{5}$ for the smallest one, and exhibits a power-law decay with its size.

Upon being switched on, the Coulomb repulsion $U$ induces a magnetization at those sites with zero confined state amplitude. Its dependence is linear in $U$, unusual scaling distinct from conventional behavior, and this is due to the mean field driven by finite magnetic moments of confined states at $U = +0$. As $U$ increases, the difference in magnetization size between these two types of sites reduces, but the magnetization size is never uniform because of quasiperiodicity of the lattice.

The magnetization profile in space is very complicated for all the $U$ values, and we tried systematic and detailed analysis for their $U$ dependence. Motivated by the previous studies for the Heisenberg model, we first tried an analysis by mapping the magnetization profile into the perpendicular space. This representation contains information on the vertex type of each site and also its local environment. The magnetization profile in the perpendicular space exhibits a clear character, i.e., fractal pattern of stars and pentagons with different sizes, and this reflects the self-similarity of the lattice structure. The profile in the perpendicular space is further analyzed by means of the singular value decomposition developed for imaging technology. We have found that while the evolution with $U$ of the magnetization profile in the perpendicular space dominantly takes place in the primary model that is quite smooth, most characteristic changes are about many modes with smaller amplitude. These would be attributed to fractal-like subpatterns in the magnetization profile.

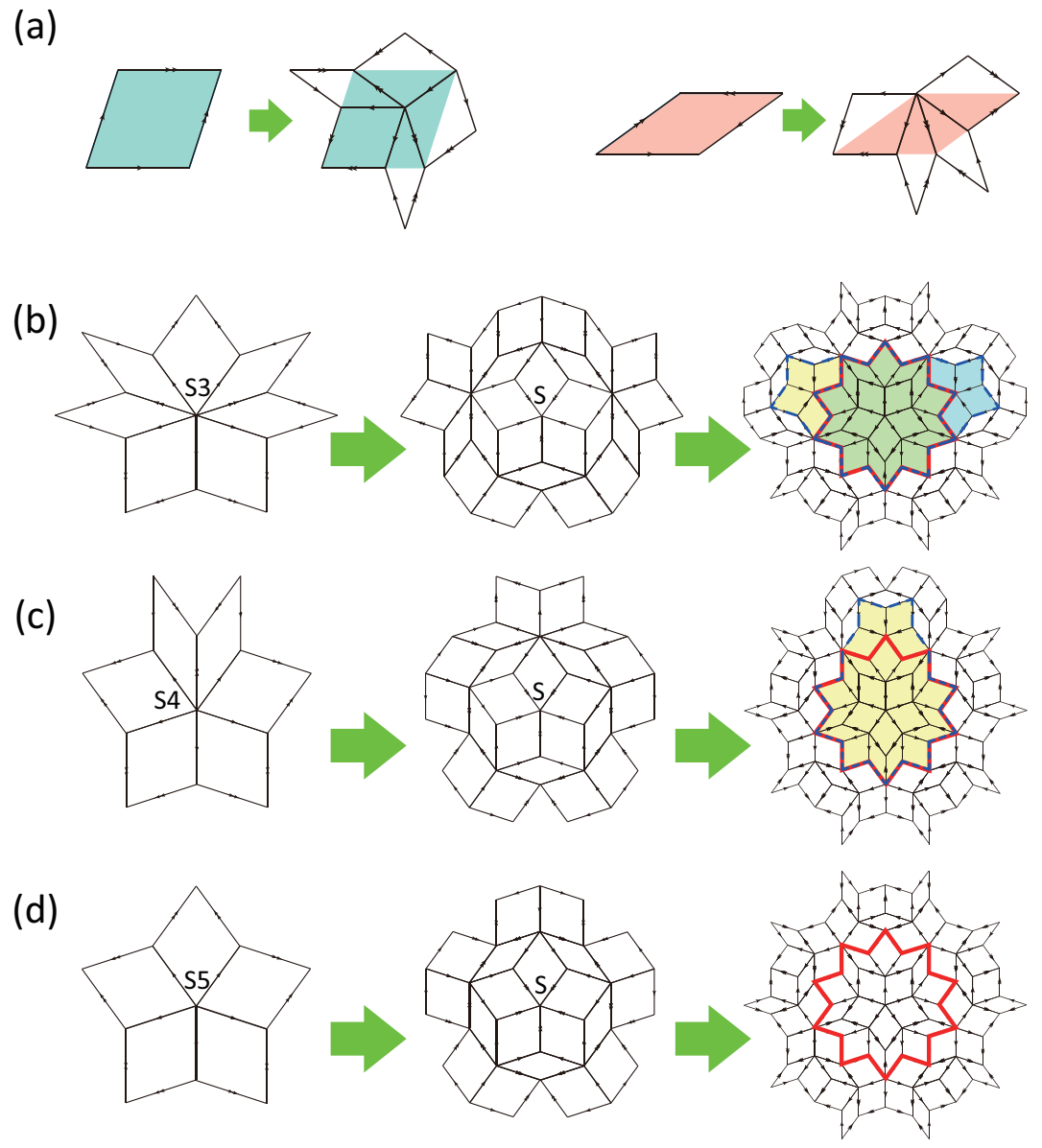

FIG. 16: Double deflation applied to (a) S3- and (b) S4-vertices. Colored parts show type-5 confined states, which are double degenerate in (a), and they overlap with a type-2 confined state shown by bold line. This overlap evidences lacking of forbidden ladder surrounding the type-2 state, and disproves that the central part is a cluster-1. Compare this with the deflation of S5-vertex shown in Fig. 3.

We also calculated the spin structure factor for several values of $U$. It has been known that the lattice structure factor has a five-fold rotation symmetry in the wave vector space, which is prohibited in periodic lattices, and a self-similar structure. Many additional peaks appear in the antiferromagnetic ordered state, and their profile depends on the $U$-value. It is interesting that the pattern has a richer texture for weaker $U$, despite weaker intensity, and this is due to magnetizations of confined states.

### Acknowledgments

The numerical calculations were mainly performed using the supercomputing systems at ISSP, the University of Tokyo. This work was supported by the Grant-in-Aid for Scientific Research from JSPS, KAKENHI Grant Number 17K05536 and 16H01066 (A.K.).

### Appendix A: Cluster-1 genesis in the deflation process

We briefly explain how the cluster-1 is generated in the deflation process. The Penrose lattice is a tiling of two types of directed rhombuses so-called “fat” and “skinny”, as shown in Fig. 3(a). Here, the single and double arrows are marked on the edges of each rhombus, following the references [27, 28]. A quasiperiodic tiling is forced by the rule that the arrow’s type and direction should match on the shared edge of each pair of adjacent rhombuses.

For generating a large lattice, the recursion relation called *deflation* is efficient, and this also leads to results about geometrical properties in an infinite-size lattice. When the deflation is applied, each rhombus divides into smaller ones as shown in Fig. 3(a), depending on if it is fat or skinny. Repeating the deflation starting from a certain initial pattern, one obtains the Penrose lattice.

Now, we demonstrate how the cluster-1’s are generated during deflation. Let us start with the pattern shown in Fig. 3 (b): five fat rhombuses centering around an S5-vertex. Applying the deflation once generates another pattern in Fig. 3(c), but this contains no cluster-1. Note that the center of the new pattern is now an S-vertex, as the arrows are pointing inwards. Applying one more deflation generates the pattern in Fig. 3 (d), and now its central part is a cluster-1, which hosts a type-2 confined state.

One should note that deflating S-vertices does not necessarily generate cluster-1’s, because the result depends on their ancestors. Both of S3- and S4-vertices are transformed to patterns around an S-vertex center. They are further transformed to larger patterns around an S5-vertex but these larger patterns do not contain a forbidden ladder encircling a cluster-1. They instead host one or two type-5 confined states as shown in See Fig. 16. Thus, double deflations do not transform S3- or S4-vertices to cluster-1’s, while deflating twice an S5-vertex always generates a cluster-1 and this is the only way of cluster-1 genesis. This proves that S5-vertices are the only source of the cluster-1 in the cluster genesis. This implies the recursion relation (2), and the relation of their densities in an infinite-size lattice,

### Appendix B: type-6 confined states

Here, we explain the mutual dependence of the two overlapping type-6 confined states shown in Fig. 5(b). In the overlapping area, there also exist other confined states of Types-1, 2, 3 and 4, and their amplitudes are shown in Fig. 17. Then, the two type-6 confined states are related to each other including the contribution of these additional states as

$$\Psi_{6B} = \Psi_{6A} + 12(\Psi_1 + \Psi_{3E}) - 5\sum_{s=A,B}\Psi_{2s} - 8\sum_{s=A,\cdots,D}\Psi_{3s} - 10\Psi_4. \quad \text{(B1)}$$

### Appendix C: Numbers of sites and confined states

In this appendix, we derive a general formula of several numbers in Table I: in the cluster-$i$, the number of the sites $N_i$, the difference in the site number between the two sublattices $\Delta N_i^{AB} = |N_i^A - N_i^B| = N_i^M - N_i^D$, and the number of S5 sites $(S5)_i$. These numbers satisfy the same recursion relation

$$X_i = 2X_{i-1} + 3X_{i-2} - 5X_{i-3}, \quad (i \geq 5) \quad \text{(C1)}$$

while their initial values depend on the numbers. Therefore, they can be represented with powers of the solutions of the

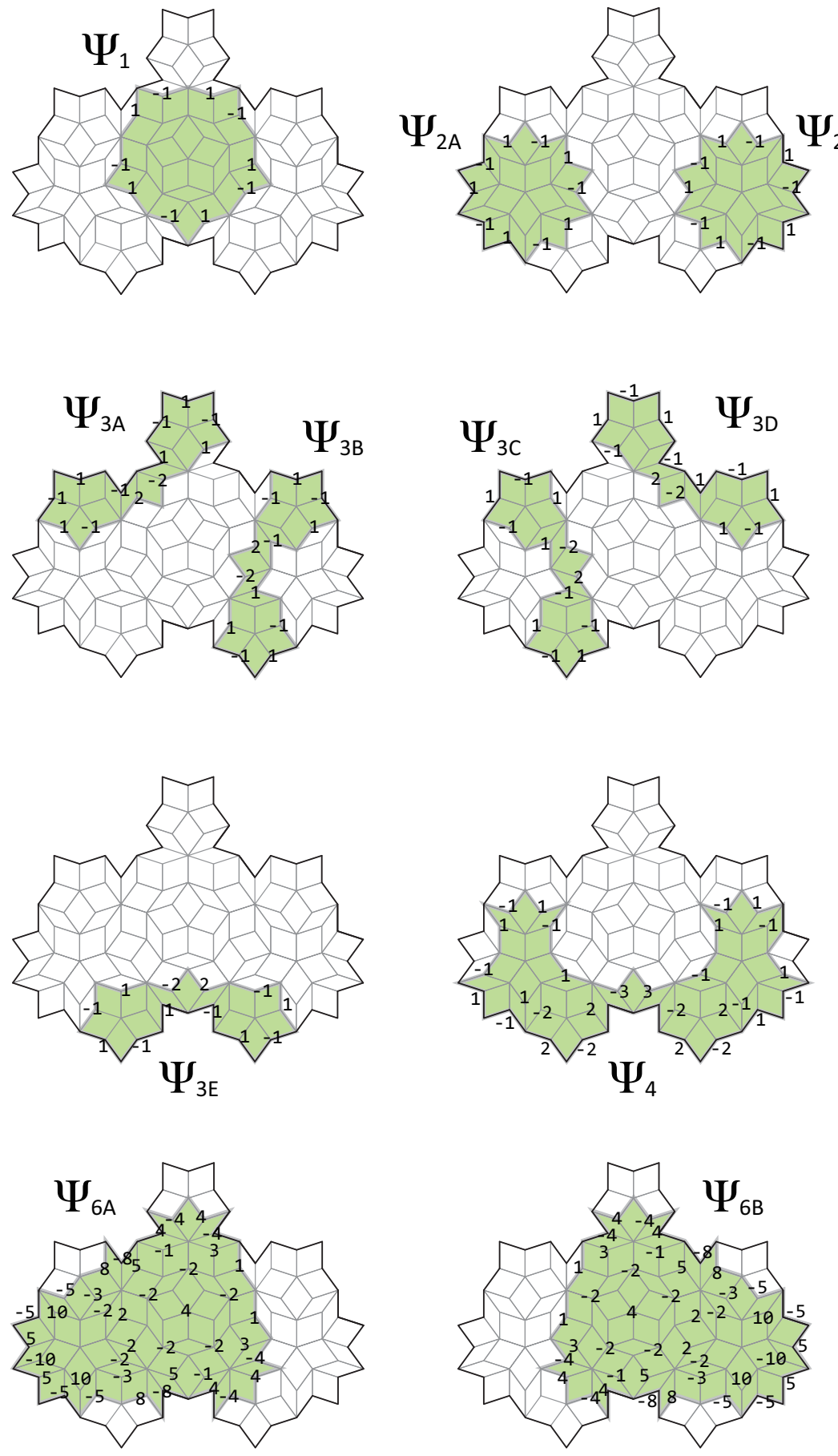

FIG. 17: Confined states appearing in the area shown in Fig. 5 (b), where two type-6 states overlap.

following characteristic equation

$$\alpha^3 = 2\alpha^2 + 3\alpha - 5. \tag{C2}$$

Its formal roots, $\alpha_l = \frac{2}{3} + \frac{2\sqrt{13}}{3}\cos\left(\frac{2l\pi}{3} - \theta_\alpha\right)$ with $\cos 3\theta_\alpha = -\frac{5}{2\sqrt{13}}$, evaluate

$$\alpha_0 = 2.37720,$$
$$\left.\begin{array}{l}\alpha_1\\ \alpha_2\end{array}\right\} = 1 - \frac{\alpha_0}{2} \pm \sqrt{\left(\frac{\alpha_0}{2} - 1\right)^2 + \frac{5}{\alpha_0}} = \left\{\begin{array}{r}1.27389\\ -1.65109\end{array}\right. \tag{C3}$$

With these values, the numbers of various sites and confined states are written as $X_i = \sum_{l=0}^{2} c_l\,\alpha_l^i$ for $i \geq 5$ with the coefficients determined from the three initial values

$$c_l = \frac{\alpha_l X_4 + \alpha_l(\alpha_l - 2)X_3 - 5X_2}{11\alpha_l^2 + 20\alpha_l - 50}. \tag{C4}$$

This leads to the results for the numbers

$$N_i = \sum_{l=0}^{2} \frac{42\alpha_l^2 + 31\alpha_l - 96}{\frac{11}{5}\alpha_l^2 + 4\alpha_l - 10}\,\alpha_l^i, \tag{C5}$$

$$\Delta N_i^{AB} = \sum_{l=0}^{2} \frac{4\alpha_l^2 + 3\alpha_l - 6}{\frac{11}{5}\alpha_l^2 + 4\alpha_l - 10}\,\alpha_l^i, \tag{C6}$$

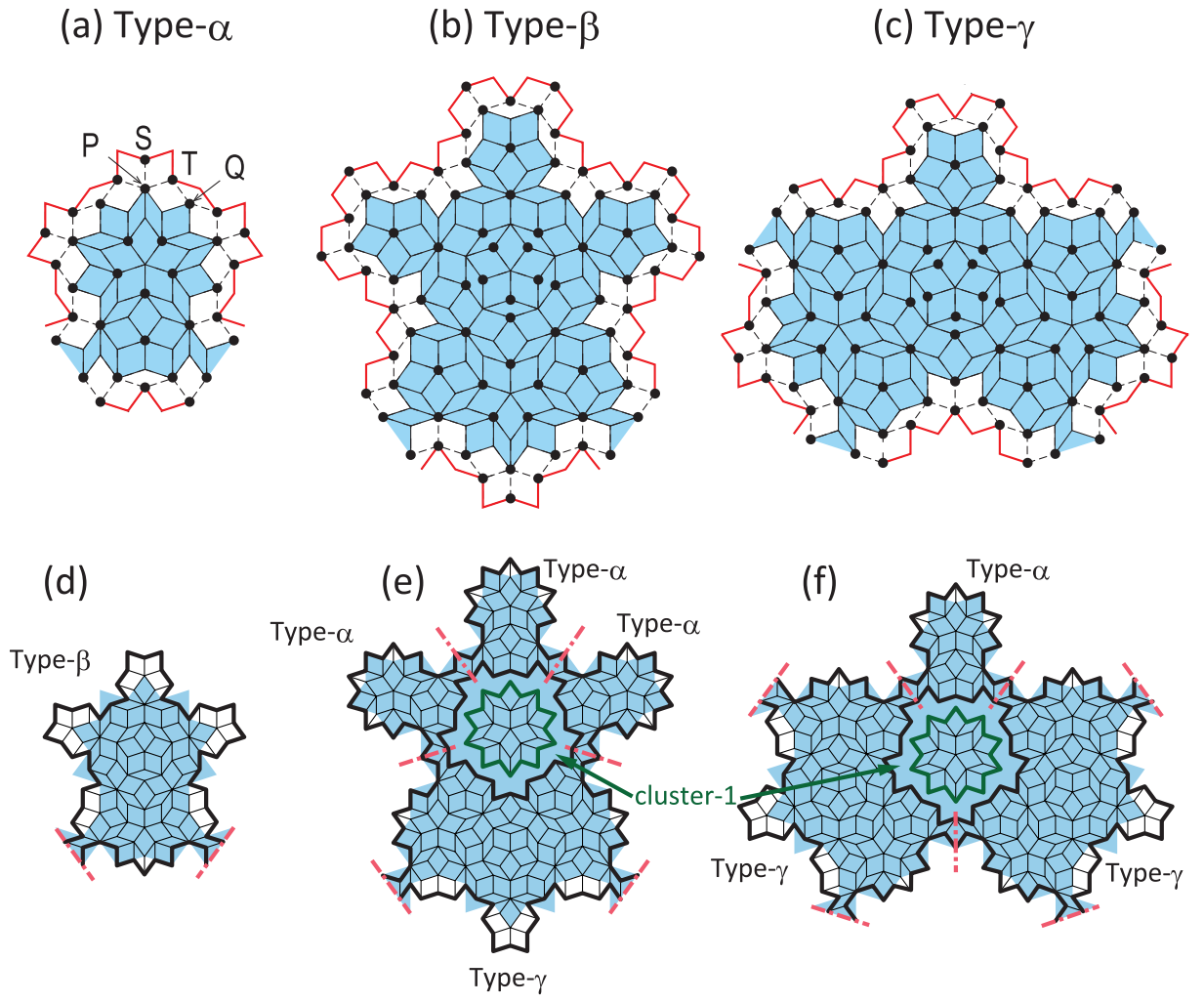

FIG. 18: (Color online) (a), (b) and (c) represent three types of subclusters. Colored region shows each subscluster. Solid circles represent the D sites. Forbidden ladders are bounded by red lines and colored region. (d), (e), and (f) represent the patterns generated by one deflation from (a), (b), and (c), respectively. The area in the initial pattern is colored blue. Dot dashed lines are boundaries for the subclusters.

$$(S5)_i = \sum_{l=0}^{2} \frac{\alpha_l^2 - \alpha_l}{\frac{11}{5}\alpha_l^2 + 4\alpha_l - 10}\,\alpha_l^i, \tag{C7}$$

We have checked the relations $\sum_{i=1}^{\infty} p_i N_i = 1$, and $\sum_{i=1}^{\infty} p_i\,(S5)_i = p_{S5}$, and these confirm that the results are correct.

The numbers of confined states in the cluster-$i$ have the following relations and thus they are calculated from $\{(S5)_{i'}\}$ already obtained in Eq. (C7):

$$N_{i,1} = N_{i-1,2} = (S5)_{i-1}, \tag{C8}$$
$$N_{i,2} = (S5)_i, \tag{C9}$$
$$N_{i,3} = 7N_{i-1,2} + 3N_{i-2,2} - 5N_{i-3,2}$$
$$= 7(S5)_{i-1} + 3(S5)_{i-2} - 5(S5)_{i-3}, \tag{C10}$$
$$N_{i,4} = N_{i-1,3}$$
$$= 7(S5)_{i-2} + 3(S5)_{i-3} - 5(S5)_{i-4}, \tag{C11}$$
$$N_{i,5} = N_{i,2} = (S5)_i \quad (i > 2), \tag{C12}$$
$$N_{i,6} = N_{i,1} = (S5)_{i-1} \quad (i > 2). \tag{C13}$$

### Appendix D: No confined states in the D sites in a cluster

Here, we show that confined state appear in only one sublattice in any cluster defined in Sec. III. Let us see the structure of the cluster-$i$ ($i \geq 3$). As is shown in Fig. 4, it is made of *subclusters* linked by narrow *bridges*. We find that all the subclusters are categorized to the three types shown in Figs. 18(a), (b) and (c). One can understand this by considering the deflation process depicted in Fig. 18. One deflation transforms a Type-$\alpha$ subcluster in the certain cluster-$i$ to a Type-$\beta$ subcluster in the generated cluster-$(i+1)$. Figure 18(e) shows that a Type-$\beta$ subcluster is transformed by deflation to a pattern containing a cluster-1. This reflects the fact that double deflations transform an S5-vertex to a cluster-1, which was discussed in Appendix A. The same figure also shows that three Type-$\alpha$ and one Type-$\gamma$ subclusters should also appear in the generated cluster. As for the Type-$\gamma$ subcluster, it is deflated to generate one Type-$\alpha$ and two Type-$\gamma$ subclusters in addition to the cluster-1, as shown in Fig. 18(f). Therefore, the set of these three types is closed under deflation and no other type of subcluster appears.

TABLE II: Profile of each subcluster-$\delta$. Its fraction is $p_\delta$, the site number $N_\delta$ and that in each sublattice $N^i_\delta$ $(i = M, D)$.

| $\delta$ | $p_\delta$ | $N_\delta$ | $N^M_\delta$ | $N^D_\delta$ |
|---|---|---|---|---|
| $\alpha$ | $\tau^{-11}$ | 42 | 23 | 19 |
| $\beta$ | $\tau^{-13}$ | 115 | 63 | 52 |
| $\gamma$ | $\tau^{-12}$ | 134 | 75 | 59 |

One should note that the cluster-3 is composed of five Type-$\alpha$ subclusters. Since all the clusters are generated from smaller clusters by deflation, this means that all of them are composed of these three kinds of subclusters. We show the profile of the subclusters in Table II. Let $N_{i,\delta}$ denote the number of Type-$\delta$ ($\delta = \alpha, \beta, \gamma$) subclusters in the cluster-$i$. They follow the recursion relation

$$\begin{bmatrix} N_{i+1,\alpha} \\ N_{i+1,\beta} \\ N_{i+1,\gamma} \end{bmatrix} = \begin{bmatrix} 0 & 3 & 1 \\ 1 & 0 & 0 \\ 0 & 1 & 2 \end{bmatrix} \begin{bmatrix} N_{i,\alpha} \\ N_{i,\beta} \\ N_{i,\gamma} \end{bmatrix}, \quad (i \geq 3) \qquad \text{(D1)}$$

and its solution reads as

$$N_{i,\alpha} = 5\sum_{l=0}^{2} \frac{3\alpha_l - 5}{11\alpha_l^2 + 20\alpha_l - 50}\alpha_l^{i-1}, \qquad \text{(D2)}$$

$$N_{i,\beta} = 5\sum_{l=0}^{2} \frac{3\alpha_l - 5}{11\alpha_l^2 + 20\alpha_l - 50}\alpha_l^{i-2}, \qquad \text{(D3)}$$

$$N_{i,\gamma} = 5\sum_{l=0}^{2} \frac{1}{11\alpha_l^2 + 20\alpha_l - 50}\alpha_l^{i}, \qquad \text{(D4)}$$

where $\alpha_l$'s are given in Eq. (C3). We have checked the relation $\sum_\delta N_\delta N_{i,\delta} = N_i$.

Next, we consider a confined state $|\psi\rangle$ in each subcluster and show that its amplitude is finite at only M sites. As explained before, when $E = 0$ eigenstates are concerned, each sublattice part is independently an eigenstate, it suffices to exclude the possibility of confined state that has finite amplitude in only the D-sublattice. Let us prove this by induction, and focus the confined-state weight at the site $P$ in Fig. 18(a). This site is inside the subcluster-$\alpha$ and in the D-sublattice. One finds $\psi(P) = 0$, since otherwise the Schrödinger equation would have led to $\psi(S) = -t\psi(P) \neq 0$ in the adjacent cluster and this contradicts the fact that these two sites are separated by a forbidden ladder shown by a white ribbon in the figure. Similarly, we can also show $\psi(Q) = 0$, since $0 = (H_0\psi)(T) = -t[\psi(P) + \psi(Q)]$. Repeating similar argument for other D-sites, we find that the confined-state weight is zero at all the D sites in the Type-$\alpha$ subcluster. We have also confirmed that this also holds in Type-$\beta$ and $\gamma$ subclusters, and this also holds in the cluster-1 and cluster-2. Since the cluster-$i$ ($i \geq 3$) is composed of three kinds of subclusters, we can conclude that no confined state appears at the D sites in any cluster.

**Appendix E: Nonlocal susceptibility**

In this appendix, we prove an important character of the nonlocal susceptibility in a bipartite lattice. The bipartiteness is an important constraint for many physical quantities of noninteracting electrons, particularly at half filling. A well-known example about the metallic case is a divergent susceptibility at the nesting vector $\mathbf{Q} = (\pi, \pi, \cdots)$ at zero temperature. The bipartiteness is also very important in nonperiodic lattices, and the corresponding property is about the sign of nonlocal susceptibility $\chi_{jj'}$:

$$\chi_{jj'} = (-1)^j(-1)^{j'}|\chi_{jj'}| \qquad \text{(E1)}$$

where $(-1)^j = +1$ if the site $j$ is in the A-sublattice and $-1$ in the B-sublattice. Therefore, $\chi_{jj'} \geq 0$ if the two sites are in the same sublattice, while $\leq 0$ if their sublattices are different. We prove this property below.

The nonlocal susceptibility at zero temperature is defined by the Kubo formula as

$$\chi_{jj'} = -\int_0^\infty d\tau \langle : c_j^\dagger(\tau)c_j(\tau) :: c_{j'}^\dagger(0)c_{j'}(0) : \rangle \qquad \text{(E2)}$$

where $\langle \cdots \rangle$ is the expectation value for the ground state and $: \cdots :$ denotes a normal order. $c(\tau)$ is the Matsubara representation of the operator. Here, the spin index is dropped for simplicity, but the charge and spin susceptibilities are both proportional to this one. Using the one-particle eigenenergies $\varepsilon_\nu$ and eigenvectors $\phi_\nu(j)$, this is represented as

$$\chi_{jj'} = \sum_{\nu,\nu'} \phi_\nu^*(j)\phi_\nu(j')\phi_{\nu'}^*(j')\phi_{\nu'}(j)\frac{f(\varepsilon_{\nu'}) - f(\varepsilon_\nu)}{\varepsilon_\nu - \varepsilon_{\nu'}}, \qquad \text{(E3)}$$

where $f(\varepsilon) = 1$ if $\varepsilon < 0$ and 0 otherwise. It is well known that the bipartness implies pairing of positive-energy states and negative energy states. Pick up any eigenstate $\phi_\nu$ with energy $\varepsilon_\nu$, and construct a new wave function $\phi'_\nu(j) \equiv (-1)^j\phi_\nu(j)$. This is also an eigenstate but its energy is $-\varepsilon_\nu$. Therefore, the special property of bipartite systems is that positive-energy eigenstates are sufficient to define any physical quantities. The result of the susceptibility is

$$\chi_{jj'} = 2(-1)^j(-1)^{j'}I_{j,j'},$$
$$I_{j,j'} \equiv \sum_{\varepsilon_\nu > 0, \varepsilon'_\nu > 0} \frac{\phi_\nu^*(j)\phi_\nu(j')\phi_{\nu'}^*(j')\phi_{\nu'}(j)}{\varepsilon_\nu + \varepsilon_{\nu'}}. \qquad \text{(E4)}$$

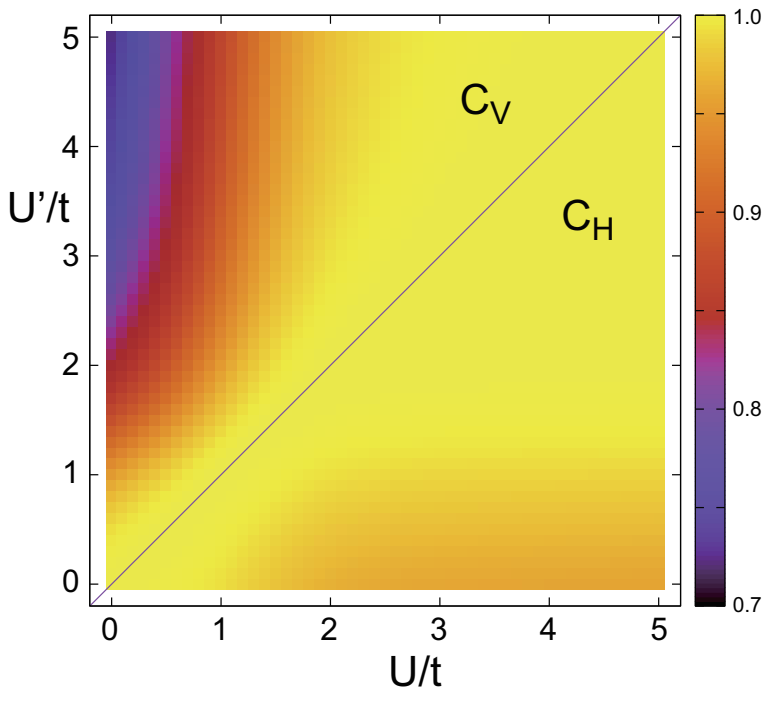

FIG. 19: Autocorrelation of the dominant mode. Lower and upper parts show $C_H(U,U')$ and $C_V(U,U')$, respectively.

Using the Laplace transformation, we can factorize $I_{j,j'}$ and prove its positivity

$$I_{j,j'} = \int_0^\infty dt \left| \sum_{\varepsilon_\nu > 0} \phi_\nu^*(j)\phi_\nu(j')e^{-\varepsilon_\nu t} \right|^2 > 0. \tag{E5}$$

This completes the proof of Eq. (E1).

### Appendix F: The detail of the mode analysis of the magnetization pattern

In this appendix, we examine the pattern $m(\tilde{x},\tilde{y})$ in perpendicular space in more detail by investigating its dominant mode and observe how the pattern changes with varying $U$. Let us define the autocorrelation function for expansion vectors as

$$C_\mu(U,U') \equiv \left[\vec{u}_\mu^{(1)}(U)\cdot\vec{u}_\mu^{(1)}(U')\right]^2, \ \ (\mu = H,V) \tag{F1}$$

and this satisfies $C_\mu(U,U) = 1$. The calculated result is shown in Fig. 19. These plots show a large but continuous crossover around $U/t \sim 1$ for the vertical direction, while small crossover around $U/t \sim 0.6$ for the horizontal direction. Expansion vectors $\vec{u}_H$ and $\vec{u}_V$ are plotted in Fig. 20 for the five largest $\lambda_\alpha$'s. Vectors for larger $\lambda_\alpha$ have a simpler coordinate dependence. For realizing a fractal-like pattern, many modes with smaller $\lambda_\alpha$ provide contributions. As for $U$ dependence, the coordinate dependence is more complicated for small $U$ and becomes smooth at large $U$.

Let us compare more quantitatively the patterns $m(\tilde{x},\tilde{y})$ in the weak and strong coupling regions. To this end, we examine how the modes $\{\vec{u}_\mu\}$ change their character between the two regions by calculating their overlaps

$$O_\mu^{\alpha\alpha'}(U,U') = \left|\vec{u}_\mu^{(\alpha)}(U)\cdot\vec{u}_\mu^{(\alpha')}(U')\right|^2, \ \ (\mu = H,V) \tag{F2}$$

Note that the previously calculated $C_\mu(U,U')$ corresponds to the largest component $\alpha = \alpha' = 1$ of this, but let us now fix $U/t = 0.0$ and $U'/t = 5.0$ and examine the dependence on modes $\alpha$ and $\alpha'$. The calculated result is shown in Fig. 21. The dominant modes with large $\lambda_\alpha$ have a large overlap between the two $U$ cases: $O^{11}$=(0.9648,0.7290), $O^{22}$=(0.7446,0.7463), $O^{33}$=(0.8530,0.8368), and $O^{44}$=(0.4913,0.7830) for $(\mu = H,V)$. This means that these modes share a quite similar character between both ends of the $U$ region. In contrast, for smaller modes (typically $\alpha > 10$), overlap matrix elements scatter over many modes in the opposite limit of $U$. This indicates that the magnetization pattern is predominantly determined by modes that vary smoothly in the projected space, and these modes change rather continuously with $U/t$ up to 5.0. However, modes with smaller $\lambda_\alpha$, which ornaments fine fractal-like subpatterns, change their character with increasing $U$.

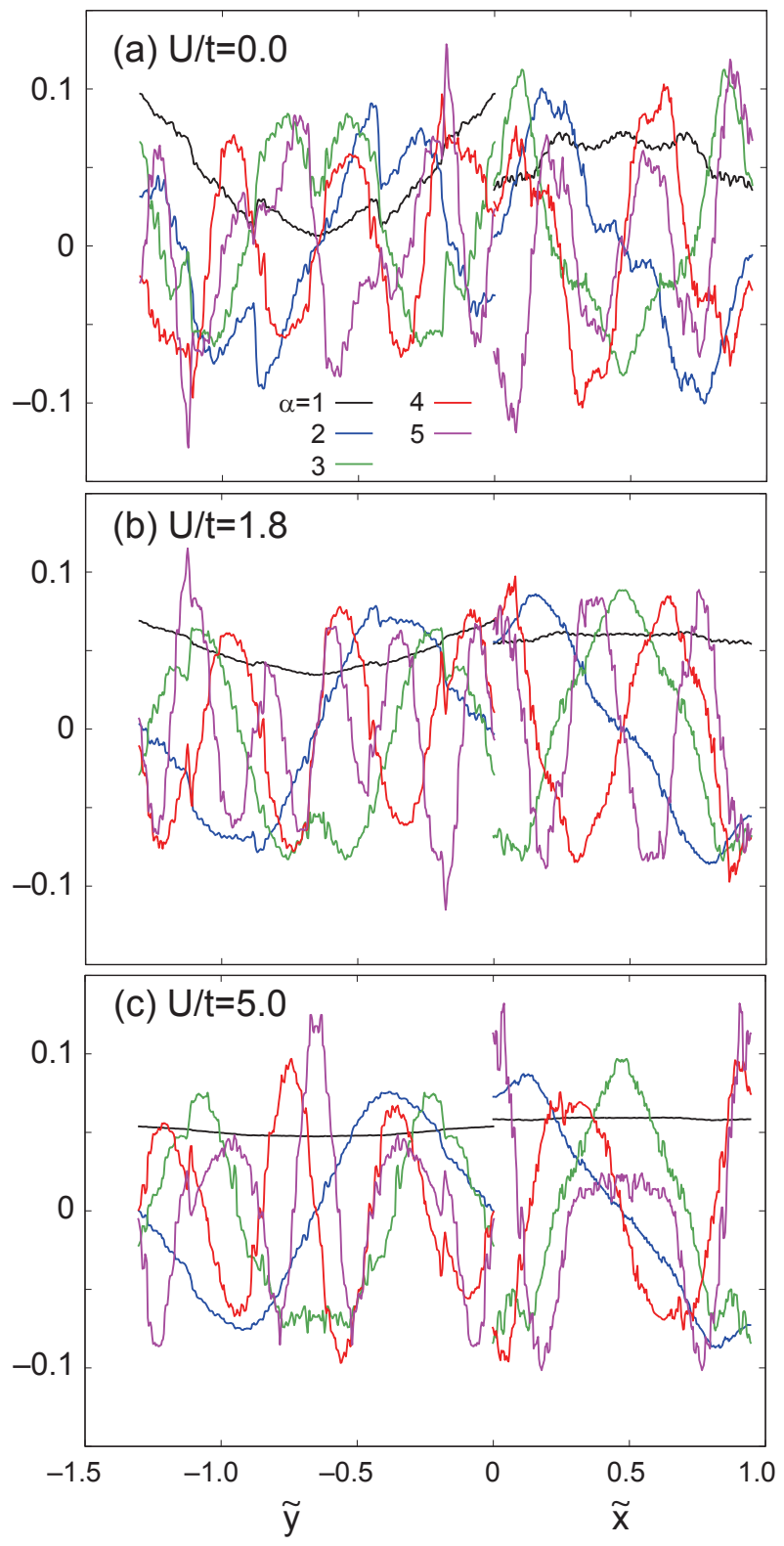

FIG. 20: Modes for the five largest singular values at three $U$ values. In each panel, the right and left parts show $\{\vec{u}_H^{(\alpha)}(\tilde{x})\}$ and $\{\vec{u}_V^{(\alpha)}(\tilde{y})\}$, respectively.

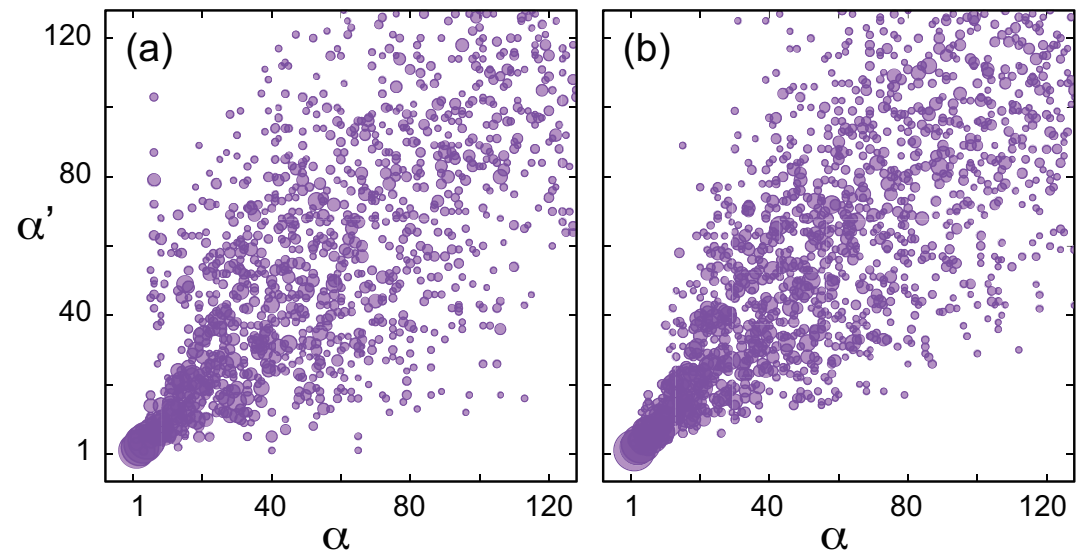

FIG. 21: Overlap of the modes determined by the singular value decomposition of $m(\tilde{x},\tilde{y})$ at $\tilde{z} = 1$. (a) $O_V^{\alpha\alpha'}(0.0,5.0)$ and (b) $O_H^{\alpha\alpha'}(0.0,5.0)$. The area of circle at $(\alpha,\alpha')$ is proportional to its value, but the data smaller than 0.02 are not shown.

---